\documentclass[a4paper,fleqn]{cas-dc}

\usepackage[numbers]{natbib}
\usepackage{amsmath,amssymb,bm}
\usepackage{graphicx}
\usepackage{booktabs}
\usepackage{cleveref}
\usepackage{xcolor}
\usepackage{siunitx}
\usepackage{subcaption}

\usepackage{multirow}%
\usepackage{amsfonts}%
\usepackage{amsthm}%
\usepackage[title]{appendix}%
\usepackage{textcomp}%
\usepackage{manyfoot}%
\usepackage{algorithm}%
\usepackage{algorithmicx}%
\usepackage{algpseudocode}%
\usepackage{listings}%
\usepackage{lineno}%
\usepackage{longtable}%
\usepackage{comment}
\usepackage{longtable}
\usepackage{setspace}
\usepackage[normalem]{ulem}

\newcommand{\RR}{\mathbb{R}}

\def\tsc#1{\csdef{#1}{\textsc{\lowercase{#1}}\xspace}}
\tsc{WGM}
\tsc{QE}

\ExplSyntaxOn
\cs_gset:Npn \__first_footerline:
  { \group_begin: \small \sffamily \__short_authors: \group_end: }
\ExplSyntaxOff
\begin{document}

\makeatletter
\def\@elsarticlemyfooter{}
\makeatother

\let\WriteBookmarks\relax
\def\floatpagepagefraction{1}
\def\textpagefraction{.001}

\shorttitle{}    

\shortauthors{}  

\title [mode = title]{Physics-informed neural networks for form-finding of unilateral membrane structures}  



%

\author[a]{Luigi Sibille}
\affiliation[a]{organization={Department of Civil and Environmental Engineering, Princeton University},
            city={Princeton},
            postcode={08544}, 
            state={NJ},
            country={United States}}
\cormark[1]
\ead{ls4272@princeton.edu}
\credit{Conceptualization, Data Curation, Formal Analysis, Investigation, Methodology, Software, Validation, Visualization, Writing - original draft, Writing - review \& editing}

\author[a]{Sigrid Adriaenssens}
\credit{Funding acquisition, Project administration, Resources, Supervision, Writing - review \& editing}

\author[c]{Carlo Olivieri}
\affiliation[c]{organization={Department of Engineering, Pegaso Telematic University},
            addressline={Centro Direzionale ISOLA F2}, 
            city={Naples},
            postcode={80143}, 
            country={Italy}} 
\credit{Conceptualization, Data curation, Investigation, Methodology, Software, Supervision, Validation, Visualization, Writing - original draft, Writing - review \& editing}

\cortext[1]{Corresponding author}

\begin{abstract}
The form-finding of unilateral membrane structures is often addressed by solving equilibrium equations using Finite Element Methods (FEMs). In this paper, Physics-Informed Neural Networks (PINNs) are investigated as an alternative, in which the equilibrium equation is enforced by minimizing its residual at collocation points through neural-network training, rather than by solving a mesh-based discretized system. This alternative is particularly suitable for form-finding problems based on Membrane Equilibrium Analysis (MEA), where the unknown is a membrane governed by a second-order elliptic Partial Differential Equation (PDE) with Dirichlet boundary conditions. This paper proposes and compares two PINN formulations: a soft-Boundary Condition (soft-BC) approach, in which the boundary conditions are enforced through a penalty term, and a hard-BC approach, in which they are satisfied by construction through distance and lift functions. The two approaches are assessed on three case studies of different geometrical complexity, involving both compression-only and tension-only stress states, as well as combined self-weight, concentrated vertical loads, and horizontal actions. The results show that both formulations provide membrane surfaces in close agreement with solutions obtained with a PDE solver based on the FEM. The hard-BC formulation yields smaller errors and a smoother spatial distribution of the residual, especially near the boundary, indicating that the exact treatment of the Dirichlet conditions has a direct influence on the overall solution accuracy. The soft-BC formulation, however, still provides structurally meaningful solutions and remains attractive when a simpler implementation is preferred and a limited relaxation of the boundary data is acceptable. Overall, the study demonstrates that PINNs are a viable numerical alternative for MEA-based form-finding and that the choice between soft and hard boundary enforcement depends on the prescribed boundary condition in the structural problem.
\end{abstract}




\maketitle

\section{Introduction}
\label{sec:introduction}

Thin shell and membrane structures owe much of their structural efficiency to the capacity to withstand loads primarily through in-plane stresses \cite{heyman1966stone, heyman1995stone}. When the stress state can be restricted to pure compression or pure tension, the resulting surface acts as a thrust membrane: a shape that conveys loads to the supports without bending, analogous in two dimensions to the classical thrust line of masonry arches \cite{heyman1977equilibrium}. Determining such shapes, the form-finding problem, is central to the conceptual design of vaults, domes, cable nets, and tensile membranes \cite{adriaenssens2014shell, melchiorre2025form, bertetto2024improved}.

Membrane Equilibrium Analysis (MEA) frames the form-finding as the solution of a second-order elliptic PDE for the thrust membrane surface \cite{olivieri2021parametric}. The approach is applicable to compression-only and tension-only membranes whose equilibrium is governed by purely compressive or purely tensile stress states \cite{angelillo2013singular}. Building on Pucher's formulation \cite{pucher1934spannungszustand}, MEA expresses the static equilibrium of a membrane shell through an Airy stress function (ASF) $\varPhi(x_1,x_2)$ whose second derivatives define the projected (Pucher) stresses \cite{angelillo2004equilibrium}. Once a concave or convex $\varPhi(x_1,x_2)$ is prescribed, guaranteeing respectively a purely compressive or purely tensile stress state, the elevation of the thrust surface $f(x_1,x_2)$ is obtained by solving the PDE with Dirichlet boundary conditions derived from the target geometry~\cite{olivieri2021parametric,olivieri2025seismic}. The formulation accommodates combined vertical and horizontal loading, where the horizontal actions enter the PDE through their corresponding load terms. In the special case of pseudo-static seismic loading, these horizontal components are taken proportional to the vertical load through load-multiplier parameters, which can be used, for example, in seismic capacity assessments of membrane structures \cite{olivieri2025seismic}. Unlike more general shell theories, such as the Naghdi model \cite{naghdi1973theory}, which describes the shell response in terms of five kinematic variables, three displacements and two rotations and accounts for membrane, bending, and shear actions, MEA is restricted to membrane equilibrium and is therefore specifically suited to the form-finding of compression-only or tension-only surfaces.

A constrained form-finding method based on the MEA was first developed in Mathematica \cite{olivieri2025seismic,olivieri2023formerly}, and has more recently been implemented in the open-source DOLFINx/FEniCSx finite-element environment \cite{olivieri2025seismic,baratta_2023_10447666}. These finite element-based implementations provide accurate PDE solutions and have proven effective for parametric membrane design. At the same time, finite-element methods remain tied to a geometric discretization of the domain and require mesh generation, matrix assembly, and the solution of a discrete algebraic system \cite{page1978finite, pingaro2026simple}. More generally, the numerical analysis of membrane structures is technically delicate: membrane formulations typically rely on differential geometry, and special care is often required to avoid locking and related discretization difficulties \cite{chapelle2011finite, Bastek2023}. Even when implementation-friendly formulations are available, the construction and use of reliable finite-element models still demand substantial expertise.

In recent years, Physics-Informed Neural Networks (PINNs) have emerged as a mesh-free alternative to compute PDEs by embedding the governing equations and boundary conditions into the loss function of a neural network \cite{raissi2019physics, son2023novel, zhou2025physics}. A PINN approximates the solution with a neural network whose parameters are optimized so that the network satisfies the differential equation at collocation points, together with the prescribed boundary conditions. Because no mesh generation or assembly step is required, PINNs can handle irregular domains and complex loading patterns with minimal additional implementation effort \cite{raissi2019physics}.

A growing body of literature has investigated both the use of PINNs for PDE solution and the improvement of their numerical behavior. The original PINN \cite{raissi2019physics} demonstrated the solution of forward and inverse PDE problems through neural networks constrained by physical laws. Since then, PINNs have been applied in several fields. For example, in fluid mechanics, PINNs have been used to infer velocity and pressure fields in complex flows and to solve forward and inverse thermal problems \cite{Cai2021}. In solid mechanics, it was further demonstrated the potential of physics-informed deep learning for inversion and surrogate modeling of elastostatic and elastoplastic problems \cite{Haghighat2021}. Within structural mechanics, PINNs were employed to shell structures \cite{Bastek2023}, showing that PINNs can solve shell equilibrium equations on curved manifolds based on the Naghdi shell model and highlighting the advantages of weak-form formulations in that context. 

Despite these promising results, the practical implementation of PINNs remains challenging. A well-known difficulty is that the loss function typically combines several loss terms, such as PDE residuals and boundary conditions through a weighted sum \cite{raissi2019physics}. When these contributions are poorly balanced, training may suffer from numerical stiffness, gradient imbalances, and poor convergence \cite{wang2021understanding, krishnapriyan2021characterizing}. This issue has motivated the development of adaptive loss-weighting strategies, including the Relative Loss Balancing with Random Lookback (ReLoBRaLo) method \cite{Bischof2025}. Other studies have addressed other well-known PINN training difficulties. Variational and weak-form formulations have been proposed to reduce the order of the derivatives appearing in the loss and to improve stability and accuracy \cite{Bastek2023, kharazmi2019variational}. Adaptive activation functions have been introduced to accelerate convergence and improve the solution approximation \cite{jagtap2020locally}. In addition, residual-based adaptive sampling strategies have been developed to place collocation points more effectively in regions where the solution is difficult to approximate \cite{Wu2023}. 

Another important aspect concerns the treatment of Dirichlet boundary conditions. In the original soft formulation \cite{raissi2019physics}, the boundary conditions are imposed through an additional penalty term in the loss, whereas in hard formulations they are satisfied exactly by embedding the boundary data typically through trial functions or distance functions \cite{Lagaris1998, berg2018unified, sukumar2022exact}. This second approach avoids the need to tune the relative weight of the boundary term and has therefore attracted considerable interest in recent PINN research.

This paper presents the use of PINNs within the MEA for form-finding of membrane surfaces. The stress state and the loading configuration are prescribed a priori, and the unknown is the membrane surface that satisfies the MEA governing equilibrium equation together with the imposed boundary conditions. In particular, two PINN configurations are proposed. The first adopts a soft enforcement of the boundary conditions, in which the Dirichlet constraint is included as an additional loss term and balanced against the PDE residual during training. The second imposes a hard enforcement of the boundary conditions, in which the solution is constructed so that the Dirichlet boundaries are satisfied exactly by means of a distance function and a lifting term. Both strategies are benchmarked against the FEniCSx solver \cite{baratta_2023_10447666} on three case studies subjected to different loading scenarios. The main objective of the paper is therefore twofold: first, to assess to what extent PINNs can provide accurate solutions for the MEA-based membrane equilibrium problem; and second, to determine the advantages and limitations of soft and hard boundary condition enforcement within the MEA.

The remainder of the paper is organized as follows. \Cref{sec:mea} reviews MEA and derives the governing PDE. \Cref{sec:pinns} presents the two PINN formulations. \Cref{sec:results} describes the three case studies and discusses the corresponding results. Finally, conclusions are drawn in \Cref{sec:conclusions}.

\section{Membrane Equilibrium Analysis}
\label{sec:mea}
This section summarizes the theoretical background of the MEA following \cite{olivieri2021parametric,angelillo2013singular,olivieri2025seismic}. The aim is to derive the second-order PDE whose solution defines the thrust membrane surface under a prescribed compression-only or tension-only stress state.

\subsection{Problem statement}
\label{sec:mea:problem}

Throughout this section, the derivatives of a generic function $\bullet(x_1,x_2)$ are defined as
\begin{equation}
\begin{aligned}
    &\bullet_{,1} = \frac{\partial \bullet}{\partial x_1}, \qquad
    \bullet_{,2} = \frac{\partial \bullet}{\partial x_2}, \\
    &\bullet_{,11} = \frac{\partial^2 \bullet}{\partial x_1^2}, \qquad
    \bullet_{,22} = \frac{\partial^2 \bullet}{\partial x_2^2}, \\
    &\bullet_{,12} = \frac{\partial^2 \bullet}{\partial x_1 \partial x_2}.
\end{aligned}
\end{equation}

\begin{figure*}[H]
  \centering
  \includegraphics[width=\textwidth]{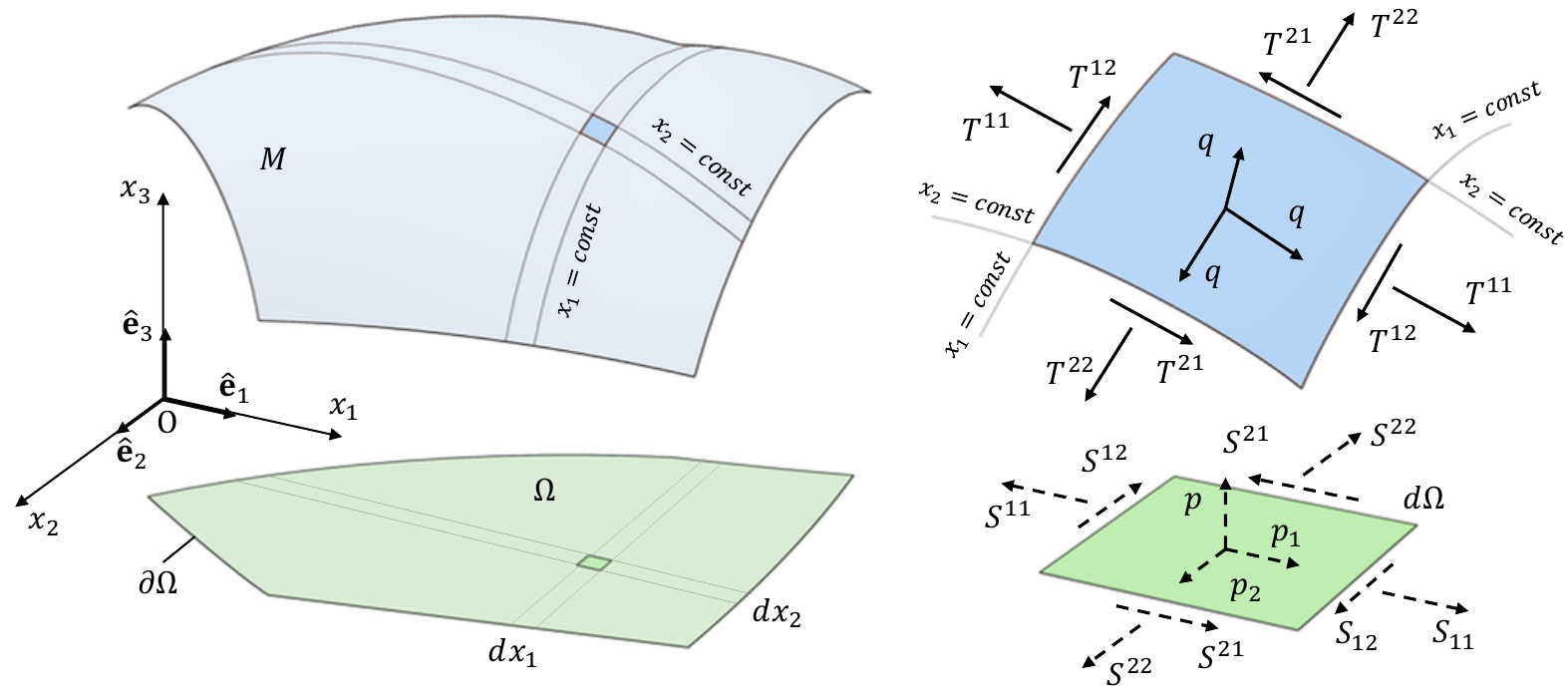}
  \caption{Geometry and loads in the MEA formulation. The membrane surface $S$ is represented above its plan projection $\Omega$. The right panel shows the projected membrane stresses and the external loads acting on an infinitesimal region of the planform.}
  \label{fig:mea_loads}
\end{figure*}

Consider a membrane surface $S$ whose plan projection lies within a bounded domain $\Omega \subset \RR^2$ with boundary $\partial \Omega$, as shown in \Cref{fig:mea_loads}. Let $\bm{x}=(x_1,x_2)\in\Omega$ denote a point in the planform. The membrane surface is described by the Monge patch
\begin{equation}
    \bm{r}(\bm{x}) = \bigl(x_1,x_2,f(x_1,x_2)\bigr),
\end{equation}
where $f:\Omega\to\RR$ represents the height of the membrane above the planform.

The membrane is subjected to external loads acting on its surface, including self-weight and possible additional actions. Within the MEA formulation, these loads are expressed in terms of their projection onto the planform $\Omega$. In particular, a vertical distributed load $q(x_1,x_2)$ is defined per unit plan area, together with horizontal load components $p_1(x_1,x_2)$ and $p_2(x_1,x_2)$, also referred to the plan projection.

As illustrated in \Cref{fig:mea_loads}, the horizontal components enter the projected equilibrium through the cumulative load resultants
\begin{equation}
  \label{eq:h1h2}
  h_1(x_1,x_2)=\int p_1\,\mathrm{d}x_1,
  \qquad
  h_2(x_1,x_2)=\int p_2\,\mathrm{d}x_2,
\end{equation}
which represent the in-plane actions associated with the horizontal loading.


\subsection{Governing PDE}
\label{sec:mea:pde}
Let $S_{11}$, $S_{22}$, and $S_{12}$ denote the projected membrane stresses acting on the planform, as illustrated in \Cref{fig:mea_loads}, obtained from the corresponding stresses on the membrane surface $T_{11}$, $T_{22}$, and $T_{12}$. In the MEA, the vertical equilibrium of the membrane reduces to a single second-order PDE for the unknown surface $f(x_{1},x_{2})$ \cite{olivieri2025seismic}:

\begin{equation}
  \label{eq:pde_general}
  S_{11} f_{,22}
  + S_{22} f_{,11}
  + 2 S_{12} f_{,12}
  - p_1 f_{,1}
  - p_2 f_{,2}
  = q.
\end{equation}

This equation expresses transverse equilibrium on the planform and, as shown in \Cref{fig:mea_loads}, couples the projected stresses with both the vertical and horizontal external actions \cite{olivieri2025seismic}.



\Cref{eq:pde_general} is posed on $\Omega$ with Dirichlet boundary conditions
\begin{equation}
  \label{eq:bc}
  f\big|_{\partial\Omega} = b(x_{1},x_{2}),
\end{equation}

where $b(x_{1},x_{2})$ is the prescribed height along the boundary.

\subsection{Airy stress function}
\label{sec:mea:airy}
In the MEA, the stress state is prescribed through an ASF $\varPhi(x_1,x_2)$. Its constant Hessian components are denoted by
\begin{equation}
  \label{eq:airy_stresses}
  N_{11} = \varPhi_{,11},
  \qquad
  N_{22} = \varPhi_{,22},
  \qquad
  N_{12} = -\varPhi_{,12}.
\end{equation}
Under horizontal loading, the projected stresses entering the governing PDE are written as
\begin{equation}
  \label{eq:airy_stresses_with_h}
  S_{11} = N_{11} - h_2,
  \qquad
  S_{22} = N_{22} - h_1,
  \qquad
  S_{12} = N_{12}.
\end{equation}
Substituting \cref{eq:airy_stresses_with_h} into \cref{eq:pde_general} gives
\begin{equation}
  \label{eq:pde_airy}
  (N_{11}-h_{2})\,f_{,22}
  + (N_{22}-h_{1})\,f_{,11}
  + 2\,N_{12}\,f_{,12}
  - p_{1}\,f_{,1}
  - p_{2}\,f_{,2}
  = q.
\end{equation}

A fundamental assumption of the MEA is that the membrane carries either compression-only or tension-only stresses over the whole domain. Therefore, the admissible stress tensor must be sign-definite everywhere: negative semi-definite for compression-only states, and positive semi-definite for tension-only states. When the stress field is represented through an ASF, this requirement translates into a curvature condition on $\varPhi$, namely on its Hessian matrix

\begin{equation}
  \label{eq:hessian_phi}
  \nabla^2 \varPhi =
  \begin{bmatrix}
    \varPhi_{,11} & \varPhi_{,12} \\
    \varPhi_{,12} & \varPhi_{,22}
  \end{bmatrix}.
\end{equation}

For compression-only membranes, $\nabla^2 \varPhi$ must be negative semi-definite throughout $\Omega$, which is equivalent to requiring $\varPhi$ to be concave. Conversely, for tension-only membranes, $\nabla^2 \varPhi$ must be positive semi-definite, so that $\varPhi$ is convex.

In simple terms, the ASF controls the sign of the projected stresses through its curvature: a concave ASF produces a compression-only state, whereas a convex ASF produces a tension-only state. This link makes the ASF a convenient means of enforcing the admissibility of the membrane stress field within the MEA, since this requirement can be built directly into the mathematical form of $\varPhi$.

In the present work, the ASF is chosen so that its Hessian is constant over the planform and yields constant projected stresses. In particular, the three scalar parameters $(\ell_1,\ell_2,\ell_3)$ are used to define the constant stress components as

\begin{equation}
\label{eq:F_components}
\begin{aligned}
  & N_{11} = \pm \, \ell_1^2, \\
  & N_{22} = \pm \, (\ell_2^2+\ell_3^2), \\
  & N_{12} = \pm \, \ell_1\ell_2
\end{aligned}
\end{equation}

where the negative sign corresponds to a concave ASF and the positive sign to a convex ASF. Since the objective of the present work is to assess the potential of PINNs within the MEA, rather than to explore highly expressive ASF parameterizations for general form-finding, a constant-Hessian ASF provides a controlled setting while retaining the mechanics of the problem. In the absence of horizontal loads, this parameterization guarantees the unilateral condition by construction, since the ASF-generated stress tensor is sign-definite for any choice of $(\ell_1,\ell_2,\ell_3)$: the negative sign yields a compression-only state, whereas the positive sign yields a tension-only state. Under horizontal loading, however, the unilateral condition must be checked on the total stress tensor. While in the compression-only state, the concavity of the ASF is always satisfied, for the tension-only case, this requires

\begin{equation}
  S_{11} \ge 0, \qquad
  S_{22} \ge 0, \qquad
  S_{11}S_{22}-S_{12}^2 \ge 0.
\end{equation}

With the adopted parameterization
\begin{equation}
  N_{11} = \ell_1^2, \qquad
  N_{22} = \ell_2^2+\ell_3^2, \qquad
  N_{12} = \ell_1\ell_2,
\end{equation}

these conditions become
\begin{equation}
\begin{aligned}
  &\ell_1^2 \ge h_2, \\
  &\ell_2^2+\ell_3^2 \ge h_1, \\
  &(\ell_1^2-h_2)(\ell_2^2+\ell_3^2-h_1)-(\ell_1\ell_2)^2 \ge 0,
\end{aligned}
\end{equation}

Therefore, in the presence of horizontal loads, the parameters $(\ell_1,\ell_2,\ell_3)$ must be chosen large enough for these inequalities to hold throughout the whole domain $\Omega$.

Since the Hessian is constant, $\varPhi$ is a quadratic polynomial centered at the planform centroid. A convenient expression is
\begin{equation}
\begin{aligned}
\varPhi(x_1,x_2)
&= \frac{1}{2}N_{22}(x_1-\bar{x}_1)^2
+ \frac{1}{2}N_{11}(x_2-\bar{x}_2)^2 \\
&\quad - N_{12}(x_1-\bar{x}_1)(x_2-\bar{x}_2) \\
&\quad + c_0 + c_1 x_1 + c_2 x_2
\end{aligned}
\end{equation}

where $c_0$, $c_1$, and $c_2$ are irrelevant to the stresses, since only second derivatives enter \cref{eq:airy_stresses}.

\section{Physics-Informed Neural Network formulations}
\label{sec:pinns}

\subsection{Network architecture}
\label{sec:pinns:arch}

\begin{figure*} [H]
    \centering
    \includegraphics[width=1\linewidth]{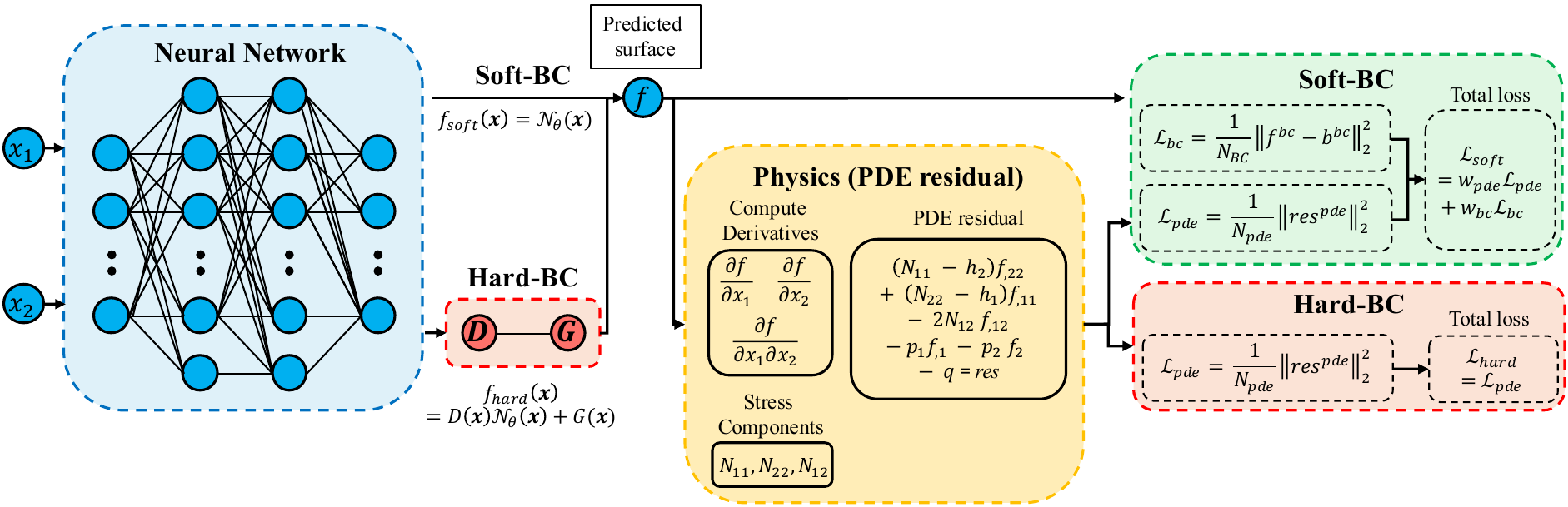}
    \caption{Overview of the proposed PINN architectures.}
    \label{fig:PINN}
\end{figure*}

Both the soft-BC and hard-BC PINN formulations approximate the height function $f(x_{1}, x_{2})$ with a fully connected feedforward neural network $\mathcal{N}_{\bm{\theta}} \colon \RR^{2} \to \RR$, where $\bm{\theta}$ collects all weights and biases. The network consists of $L$ hidden layers, each of width $W$ using GELU \cite{hendrycks2016gaussian} as activation function. The input layer receives the plan coordinates $(x_{1}, x_{2})$ and the output layer produces a single scalar. An overview of the proposed PINN formulations is shown in \Cref{fig:PINN}. 

Since \cref{eq:pde_general} involves second-order derivatives of $f(x_{1}, x_{2})$, the activation function must be at least $C^{2}$ in order to ensure a sufficiently smooth PDE residual. For this reason, only smooth activations were considered. Among the examined activation functions, i.e. tanh \cite{goodfellow2016deep}, sigmoid \cite{goodfellow2016deep}, GELU \cite{hendrycks2016gaussian}, and SiLU/Swish \cite{ramachandran2017searching,elfwing2018sigmoid}, GELU provided the best performance, and was therefore adopted in all subsequent analyses.

Different depths and widths were explored in preliminary convergence studies, and the final architectures were chosen as those providing the lowest total loss. The resulting configurations are:

\begin{itemize}
  \item \textbf{Soft-BC}: $L=5$ hidden layers with $W=128$ neurons each ($5\times128$);
  \item \textbf{Hard-BC}: $L=4$ hidden layers with $W=256$ neurons each ($4\times256$).
\end{itemize}

\subsection{Soft-BC formulation}
\label{sec:pinns:soft}

In the soft-BC approach the network output directly approximates the membrane height:
\begin{equation}
  \label{eq:soft_ansatz}
  f_{\mathrm{soft}}(\bm{x}) = \mathcal{N}_{\bm{\theta}}(\bm{x}).
\end{equation}

Boundary conditions are enforced approximately through a penalty term in the loss function.  The total loss is
\begin{equation}
  \label{eq:soft_loss}
  \mathcal{L}_{\mathrm{soft}}
  = w_{\mathrm{pde}}\,\mathcal{L}_{\mathrm{pde}}
  + w_{\mathrm{bc}}\,\mathcal{L}_{\mathrm{bc}},
\end{equation}

where:
\begin{itemize}
  \item The PDE residual loss is the mean squared residual of \cref{eq:pde_general} evaluated at $N_{\mathrm{pde}}$ interior collocation points $\{\bm{x}_{i}^{\mathrm{pde}}\}$:
    \begin{equation}
      \label{eq:pde_loss}
      \mathcal{L}_{\mathrm{pde}}
      = \frac{1}{N_{\mathrm{pde}}}\,
        \bigl\|res^{\mathrm{pde}}\bigr\|_2^2,
    \end{equation}
  \item The boundary loss is the mean squared error on $N_{\mathrm{bc}}$ boundary collocation points:
    \begin{equation}
      \label{eq:bc_loss}
      \mathcal{L}_{\mathrm{bc}}
      = \frac{1}{N_{\mathrm{bc}}}\,
        \bigl\|\bm{f}^{\mathrm{bc}}-\bm{b}^{\mathrm{bc}}\bigr\|_2^2,
    \end{equation}
    where
    \begin{align*}
      \bm{f}^{\mathrm{bc}}
      =
      \bigl[
      f(\bm{x}_1^{\mathrm{bc}}),\dots,f(\bm{x}_{N_{\mathrm{bc}}}^{\mathrm{bc}})
      \bigr]^{\top}, \\
      \bm{b}^{\mathrm{bc}}
      =
      \bigl[
      b(\bm{x}_1^{\mathrm{bc}}),\dots,b(\bm{x}_{N_{\mathrm{bc}}}^{\mathrm{bc}})
      \bigr]^{\top}.
    \end{align*}
\end{itemize}
The weights $w_{\mathrm{pde}}$ and $w_{\mathrm{bc}}$ in \cref{eq:soft_loss} are adapted during training using the ReLoBRaLo algorithm \cite{Bischof2025}. At each epoch, ReLoBRaLo computes tentative weights proportional to the exponential of the relative loss improvement over a randomly selected past epoch.  The result is a convex combination of the previous weights and the tentative ones, controlled by a balance parameter $\rho$. This mechanism prevents any single loss component from dominating the gradient and has been shown to improve convergence in multi-objective PINN training.

\subsection{Hard-BC formulation}
\label{sec:pinns:hard}
In the hard-BC approach the trial function is constructed so that the Dirichlet condition \cref{eq:bc} is satisfied identically:
\begin{equation}
  \label{eq:hard_ansatz}
  f_{\mathrm{hard}}(\bm{x})
  = D(\bm{x})\,\mathcal{N}_{\bm{\theta}}(\bm{x}) + G(\bm{x})
\end{equation}

where:
\begin{itemize}
  \item $D(\bm{x})$ is a smooth, non-negative distance-like function that vanishes on $\partial\Omega$ and is strictly positive in the interior.
  \item $G(\bm{x})$ is a harmonic lift of the boundary data: a function satisfying $G\big|_{\partial\Omega} = b$ and $\Delta G = 0$ in $\Omega$. 
\end{itemize}

Because $D$ vanishes on $\partial\Omega$, the trial function \cref{eq:hard_ansatz} reduces to $G(\bm{x})$ on the boundary regardless of $\mathcal{N}_{\bm{\theta}}$. The boundary condition is therefore satisfied exactly, and no penalty term is needed. The loss reduces to the PDE residual alone:

\begin{equation}
  \label{eq:hard_loss}
  \mathcal{L}_{\mathrm{hard}} = \mathcal{L}_{\mathrm{pde}}.
\end{equation}

\subsection{Training protocol}
\label{sec:pinns:training}

Both formulations follow a two-stage optimization protocol:
\begin{enumerate}
  \item \textbf{Stage~1 --- Adam.}
    The network is trained for $E_{\mathrm{Adam}}$ epochs using the Adam optimiser~\cite{kingma2014adam} with a constant learning rate $\eta_{\mathrm{Adam}}$.

  \item \textbf{Stage~2 --- L-BFGS.}
    Starting from the best checkpoint of Stage~1, the optimization switches to L-BFGS with a strong Wolfe line search~\cite{liu1989limited}.
\end{enumerate}

\subsection{Collocation resampling}
\label{sec:pinns:resampling}

A well-known issue in PINNs is that training on a fixed set of collocation points can lead to the network overfitting the residual at those specific locations while ignoring the PDE elsewhere. Following the practice introduced in~\cite{daw2022rethinking, daw2022mitigating}, collocation points are periodically resampled during training.

At every $K_{\mathrm{resample}}$ epochs during the Adam and L-BFGS stages, a fresh set of $N_{\mathrm{pde}}$ interior collocation points is drawn uniformly over the domain~$\Omega$. 

Boundary collocation points are likewise resampled at the same frequency.  The base set of $N_{\mathrm{bc}}$ points is drawn uniformly along the boundary (uniform arc-length for curved domains). For the 3-leg and 4-leg geometries, an additional set of $N_{\mathrm{bc,curv}}$ enrichment points is sampled from a curvature-weighted density:
\begin{equation}
  \label{eq:curv_density}
  p(\theta) \;\propto\; 1 + \frac{|\kappa(\theta)|}{\max_{\theta}|\kappa(\theta)|},
\end{equation}
where $\kappa(\theta) = \mathrm{d}^{2}g / \mathrm{d}\theta^{2}$ is the second derivative of the boundary profile. This density concentrates additional samples near the leg tips, where the boundary data has high curvature and is therefore hardest to represent.  The combined set of $N_{\mathrm{bc}} + N_{\mathrm{bc,curv}}$ points constitutes the boundary collocation cloud for each epoch.

\subsection{Hyperparameter summary}
\label{sec:pinns:hyperparams}

\Cref{tab:hyperparams} lists all hyperparameters used across the three case studies. With the exception of the network depth and width, which differ between the soft-BC and hard-BC formulations as noted in \cref{sec:pinns:arch}, the same values are used for all three domains.

\begin{table}[H]
  \centering
  \caption{PINN hyperparameters (common to all three domains unless
           noted otherwise).}
  \label{tab:hyperparams}
  \footnotesize
  \begin{tabular}{lll}
    \toprule
    \textbf{Parameter} & \textbf{Symbol / key} & \textbf{Value} \\
    \midrule
    \multicolumn{3}{l}{\emph{Architecture}} \\
    Hidden layers / width (soft-BC) & $L \times W$ & $5 \times 128$ \\
    Hidden layers / width (hard-BC) & $L \times W$ & $4 \times 256$ \\
    Activation function & --- & GELU \\
    \midrule
    \multicolumn{3}{l}{\emph{Stage~1 — Adam}} \\
    Learning rate & $\eta_{\mathrm{Adam}}$ & $10^{-3}$ \\
    Number of epochs & $E_{\mathrm{Adam}}$ & $30\,000$ \\
    \midrule
    \multicolumn{3}{l}{\emph{Stage~2 — L-BFGS}} \\
    Total steps & $E_{\mathrm{LBFGS}}$ & $10\,000$ \\
    Line search & --- & strong Wolfe \\
    Step size & $\eta_{\mathrm{LBFGS}}$ & 1.0 \\
    \midrule
    \multicolumn{3}{l}{\emph{Collocation}} \\
    Interior PDE points & $N_{\mathrm{pde}}$ & $16\,384$ \\
    Boundary points (base) & $N_{\mathrm{bc}}$ & $1\,024$ \\
    Boundary enrichment (curv.) & $N_{\mathrm{bc,curv}}$ & $1\,024$\,\textsuperscript{a} \\
    Resampling period & $K_{\mathrm{resample}}$ & 10 epochs \\
    \midrule
    \multicolumn{3}{l}{\emph{ReLoBRaLo (soft-BC only)}} \\
    EMA coefficient & $\alpha$ & 0.999 \\
    Bernoulli anchor probability & $\rho$ & 0.8 \\
    Softmax temperature & $\tau$ & 2.0 \\
    \bottomrule
    \multicolumn{3}{l}{\textsuperscript{a}\,3-leg and 4-leg domain only.} \\
  \end{tabular}
\end{table}

\section{Results}
\label{sec:results}
\begin{figure*}[H]
  \centering
  \includegraphics[width=\textwidth]{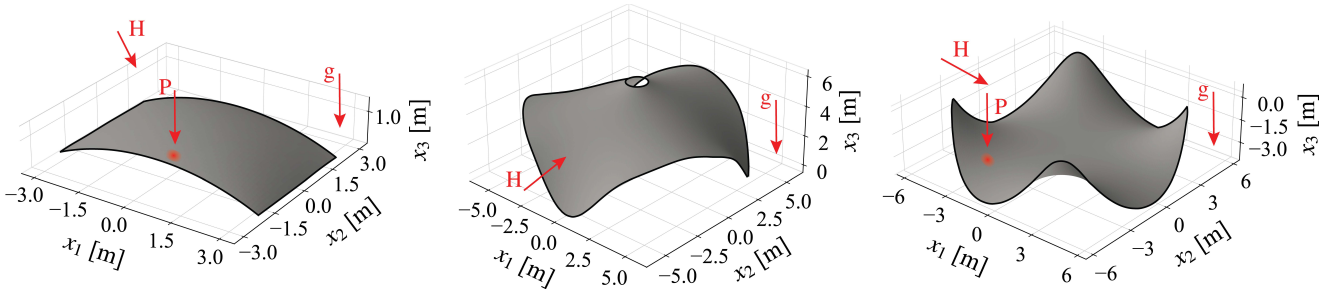}
    \caption{Considered case studies together with their boundary conditions and load configurations. From left to right: rectangular domain, the three-legged domain, and the four-legged domain. The supports are highlighted in black, while the self-weight $q$ load, the applied vertical point load $P$, and the horizontal action $H$  are indicated in red.}  \label{fig:geometries}
\end{figure*}

The PINN architectures described in \Cref{sec:pinns} are applied to three case studies with different geometrical complexity. For each case, the finite-element reference solution is computed with FEniCSx, and the PINN solutions are evaluated on the same set of nodal points. The membranes of the three case studies, together with the corresponding boundary conditions and applied loading, are shown in \Cref{fig:geometries}. The membrane surfaces shown in the figure are membranes obtained by solving the MEA governing PDE with FEniCSx for the prescribed boundary conditions, the selected Airy stress parameters, and the self-weight of the membranes only. The additional concentrated and horizontal loads are not included in the PDE, but they are only indicated schematically to clarify the loading setup adopted in each case study. The corresponding solutions under the complete loading actions are presented in \Cref{sec:results:soft,sec:results:hard}.

\subsection{Case studies}
\label{sec:results:case_studies}

The first case considers a rectangular plan domain of dimensions $l=\SI{13}{m}$ and $b=\SI{8}{m}$. The membrane is assumed to have thickness $t=\SI{0.1}{m}$ and is subjected to self-weight corresponding to a material specific weight $\rho=\SI{18}{kN/m^3}$, together with a concentrated load of \SI{5}{kN} applied at $(-1.5,\,0)$\,m. Since the MEA is formulated in terms of loads distributed per unit plan area, a purely concentrated point load would enter the problem as a singular term and is therefore regularized. Following the approach adopted in \cite{olivieri2023continuous}, the concentrated load is represented by a Gaussian distribution with spread $\sigma_{\mathrm{load}}=\SI{0.5}{m}$. Horizontal load components $p_1(\bm{x})$ and $p_2(\bm{x})$ are then introduced through a pseudo-static assumption, by taking them proportional to the total vertical load $q(\bm{x})$, which includes both self-weight and the regularized concentrated load. In particular, a horizontal action is applied along the diagonal direction with coefficient $\alpha_h=0.5$, so that the horizontal load components are defined as $p_1=\lambda_1 q$ and $p_2=\lambda_2 q$, where $\lambda_1=\alpha_h \cos\theta$ and $\lambda_2=\alpha_h \sin\theta$, with $\theta$ denoting the direction of the applied horizontal action. The Airy parameters $(\ell_{1},\ell_{2},\ell_{3})=(2,0,2)$ define a constant compressive stress state, with $N_{11}=N_{22}=-4$ and $N_{12}=0$.

The second case considers a three-legged annular domain inscribed on a disk with outer radius $R_{\mathrm{out}} \approx \SI{6.0}{m}$ and inner radius $R_{\mathrm{in}} \approx \SI{0.6}{m}$. Also in this case, the membrane thickness is taken as $t=\SI{0.1}{m}$. The membrane is subjected to self-weight with $\rho=\SI{18}{kN/m^3}$ and to a pseudo-static horizontal action applied with $\alpha_h=0.3$ acting along the $x_2$ direction ($\theta=\pi/2\,\mathrm{rad}$). No concentrated load is applied. The Airy parameters are $(\ell_{1},\ell_{2},\ell_{3})=(3,0,3)$, which yield the constant stress components $N_{11}=N_{22}=-9$ and $N_{12}=0$.

The third case uses a four-legged membrane inscribed on a disk of radius $R\approx \SI{6.0}{m}$. The membrane thickness is again set equal to $t=\SI{0.1}{m}$. The applied loads include self-weight with $\rho=\SI{10}{kN/m^3}$, a concentrated load of \SI{15}{kN} applied at $(-2.5,\,-2.5)$\,m, and a pseudo-static horizontal action with $\alpha_h=0.5$ acting along the $x_1$ direction ($\theta=0\,\mathrm{rad}$). In contrast with the other two examples, this case is formulated as a tension-only problem. Accordingly, the Airy parameters $(\ell_{1},\ell_{2},\ell_{3})=(5,0,5)$ define a convex ASF and constant tensile stress components, with $N_{11}=25$, $N_{22}=25$, and $N_{12}=0$.

\begin{figure*}[H]
  \centering
  \includegraphics[width=\textwidth]{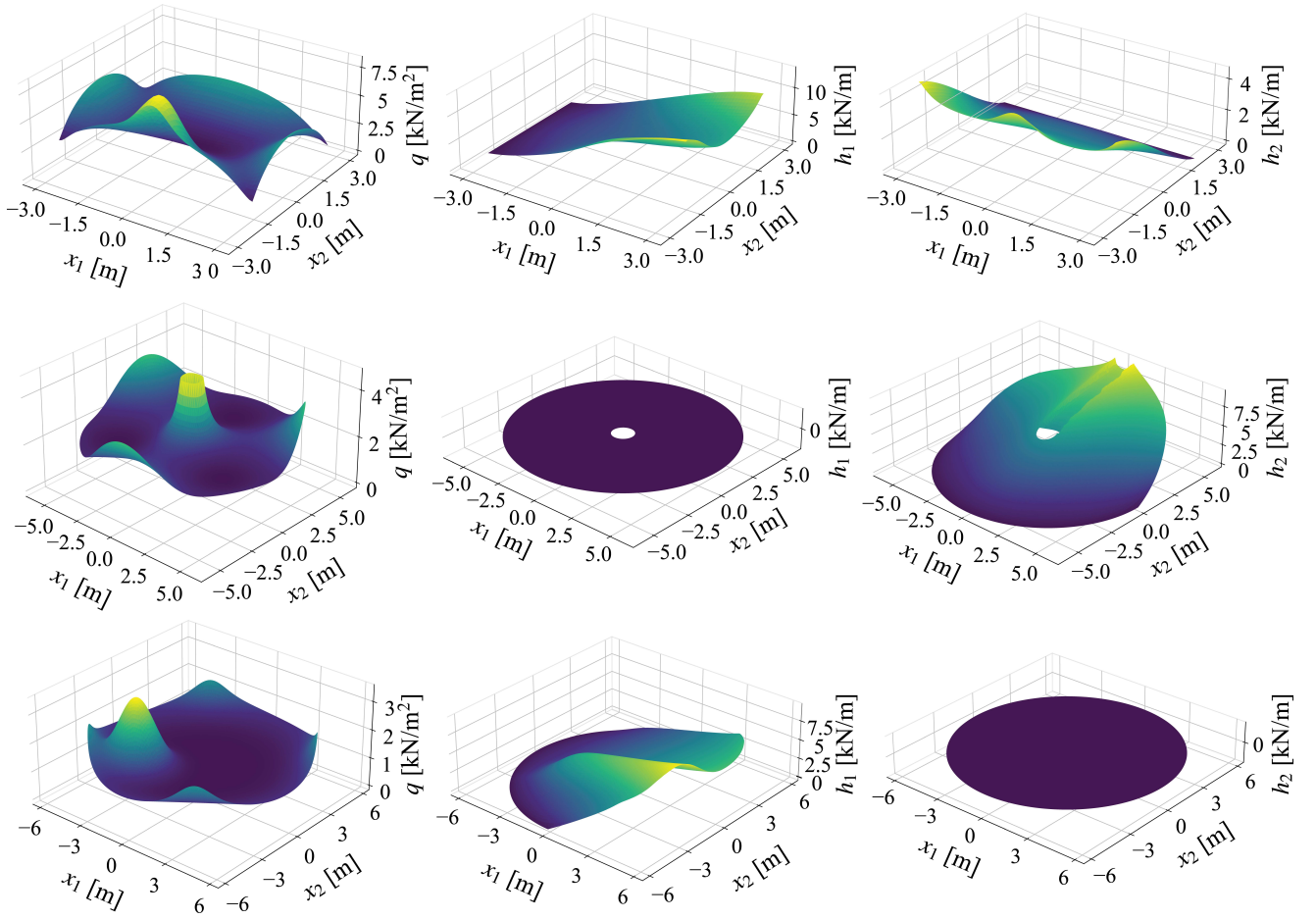}
  \caption{Vertical load $q$, cumulative horizontal load $h_{1}$, and cumulative horizontal load $h_{2}$ entering the governing PDE. From left to right, the columns correspond to the rectangular domain, the three-legged domain, and the four-legged domain}
  \label{fig:applied_loads}
\end{figure*}

For completeness, the applied loads are reported in \Cref{fig:applied_loads}, whereas the ASF parameters, horizontal load coefficients, and main geometric quantities are summarized in \Cref{tab:case_params}. The ASF surfaces associated with the selected parameters, together with the corresponding principal membrane stresses and principal stress directions, are reported in Appendix \ref{app:airy_stresses}.

\begin{table*} [H]
  \centering
  \caption{Summary of the input parameters for the three case studies.}
  \label{tab:case_params}
  \begin{tabular}{@{}lccc@{}}
    \toprule
    Parameter & Rectangular & Three-legged & Four-legged \\
    \midrule
    Domain type
      & Rectangle & Annular disk & Disk \\
    Characteristic size
      & $13\times8$\,m & $R_{\mathrm{in}}\approx0.6$\,m, $R_{\mathrm{out}}\approx6.0$\,m & $R\approx6.0$\,m \\
    Thickness $t$
      & \SI{0.1}{m} & \SI{0.1}{m} & \SI{0.1}{m} \\
    $(\ell_{1},\ell_{2},\ell_{3})$
      & $(2,\,0,\,2)$ & $(3,\,0,\,3)$ & $(5,\,0,\,5)$ \\
    $\rho$
      & \SI{18}{kN/m^3} & \SI{18}{kN/m^3} & \SI{10}{kN/m^3} \\
    $\alpha_{h}$
      & 0.5 & 0.3 & 0.5 \\
    $\theta$
      & Diagonal & $90^{\circ}$ & $0^{\circ}$ \\
    Concentrated load
      & \SI{5}{kN} at $(-1.5,\,0)$ & --- & \SI{15}{kN} at $(-2.5,\,-2.5)$ \\
    Stress state
      & Compression-only & Compression-only & Tension-only \\
    \bottomrule
  \end{tabular}
\end{table*}

\subsection{Soft-BC formulation}
\label{sec:results:soft}

\begin{figure*}[H]
  \centering
  \includegraphics[width=\textwidth]{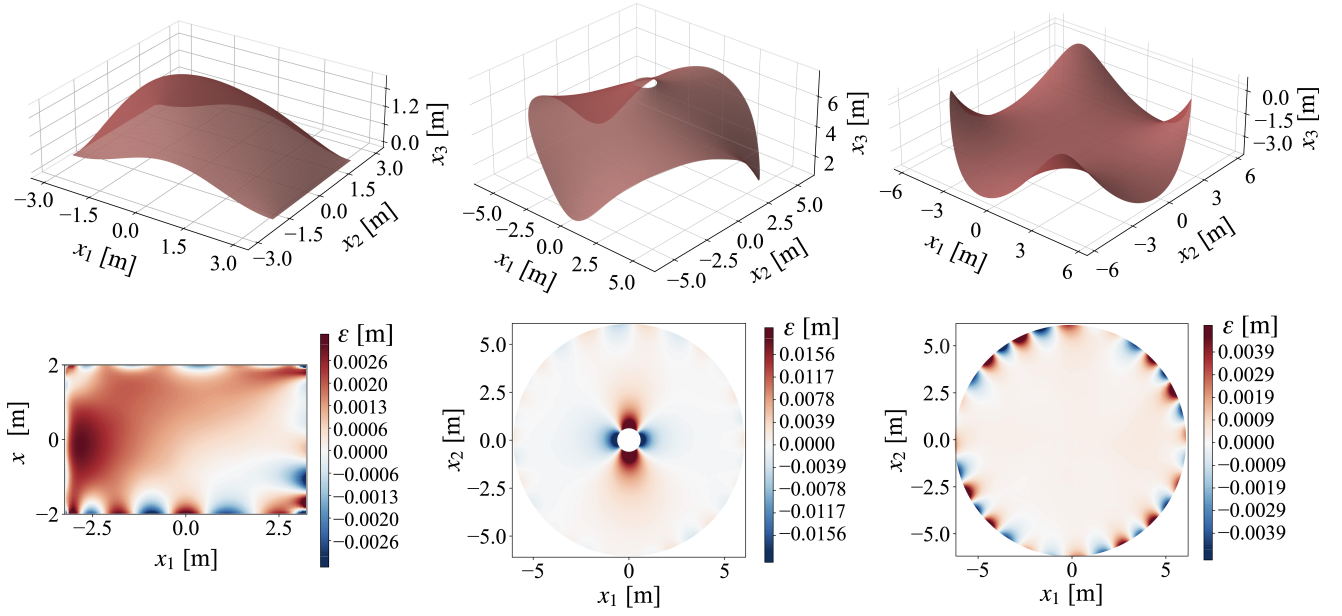}
  \caption{The top row compares the membrane surfaces obtained with the Soft-BC PINN solution (red) and with FEniCSx (gray). The bottom row shows the pointwise difference between the two solutions, computed as $f_{\mathrm{PINN}}-f_{\mathrm{FEniCSx}}$. From left to right, the columns correspond to the rectangular domain, the three-legged domain, and the four-legged domain.}
  \label{fig:diff_soft}
\end{figure*}

\Cref{fig:diff_soft} compares the membrane surfaces obtained with the soft-BC formulation (red membrane) and the FEniCSx reference (gray membrane) for the three case studies. A close agreement is observed in all cases, indicating that the neural network reproduces the reference membranes with high accuracy. The bottom row shows the pointwise difference $f_{\mathrm{PINN}}-f_{\mathrm{FEniCSx}}$. The Root Mean Squared Error (RMSE) is equal to $1.45\times10^{-3}$\,m for the rectangular domain, $3.82\times10^{-3}$\,m for the three-legged domain, and $1.00\times10^{-3}$\,m for the four-legged domain. The relative $L^{2}$ errors, i.e. the RMSE normalized by the Root Mean Square (RMS) of the reference membrane obtained with FEniCSx, is below $0.12\,\%$ in all cases. The maximum absolute error is equal to $8.47\times10^{-3}$\,m for the rectangular domain, $3.38\times10^{-2}$\,m for the three-legged domain, and $9.69\times10^{-3}$\,m for the four-legged domain. The errors are summarized in \Cref{tab:results}. In all three cases, the largest errors are concentrated near the boundary, particularly where the cumulative load terms attain their lowest values. Localized errors along the boundary are consistent with the penalty-based enforcement of the Dirichlet boundary conditions, which allows small residual mismatches at the boundary that then propagate into the interior through the elliptic character of the governing PDE.

It is worth noting that, before adopting ReLoBRaLo (\Cref{sec:pinns:soft}), a separate benchmark study was conducted to compare different loss-weighting strategies, namely fixed weights, Wang weighting \cite{wang2021understanding}, SoftAdapt \cite{heydari2019softadapt}, and ReLoBRaLo \cite{Bischof2025}. Among the strategies examined, ReLoBRaLo provided the best overall performance for the present problem and was therefore selected for the soft-BC formulation.

\subsection{Hard-BC formulation}
\label{sec:results:hard}

\begin{figure*}[H]
  \centering
  \includegraphics[width=\textwidth]{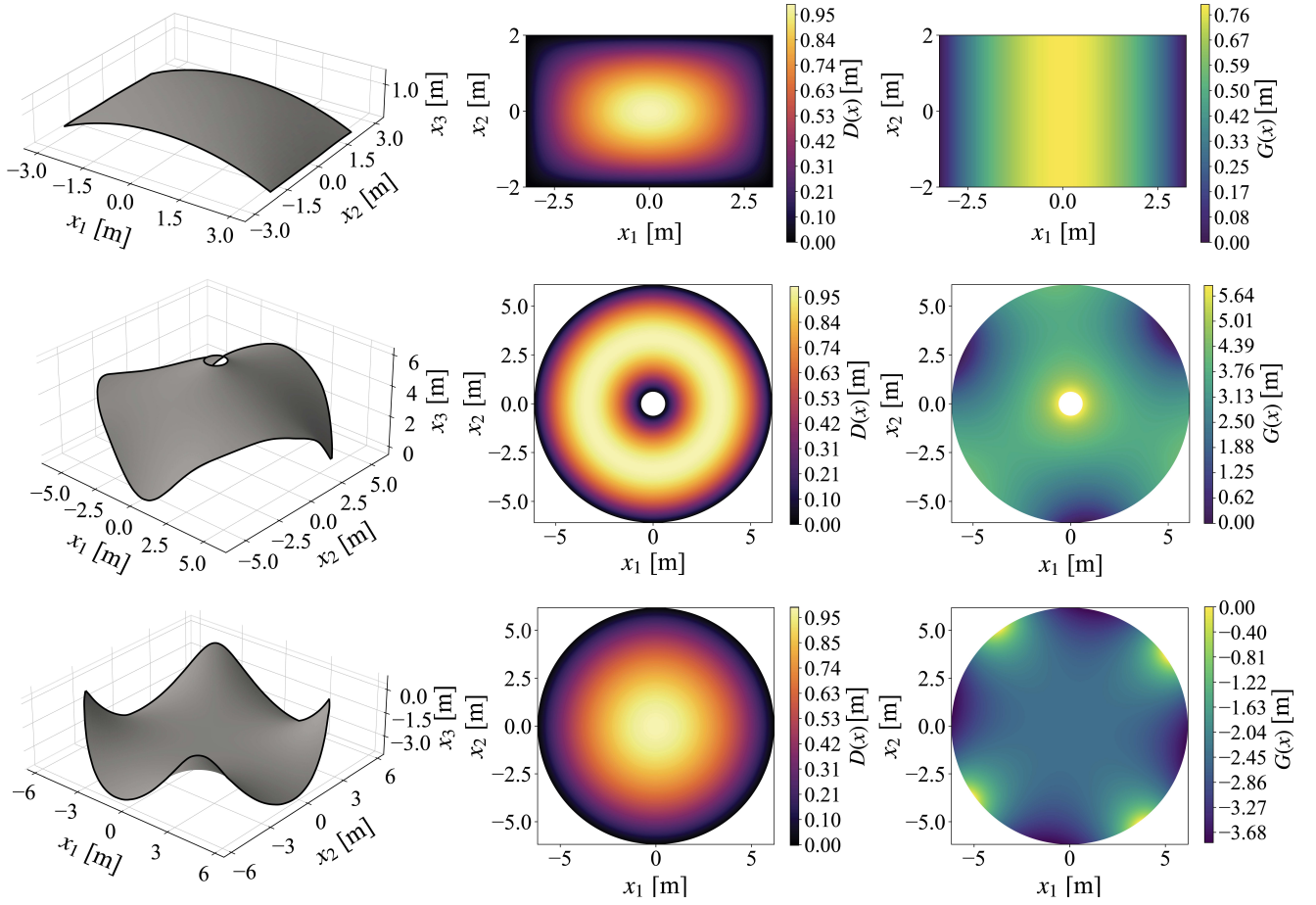}
  \caption{Distance and lifting functions adopted in the hard-BC formulation for the three case studies. The rows correspond, from top to bottom, to the rectangular domain, the three-legged domain, and the four-legged domain. The columns show, from left to right, the reference membrane geometry, the distance function $D(\bm{x})$, and the lifting function $G(\bm{x})$.}
  \label{fig:distance_lift}
\end{figure*}

The hard-BC formulation enforces the Dirichlet condition exactly through the distance function $f(\bm{x}) = D(\bm{x})\!\cdot\!\mathcal{N}+G(\bm{x})$. \Cref{fig:distance_lift} shows the distance function $D(\bm{x})$ and the lifting function $G(\bm{x})$ applied to the three case studies, while their explicit expressions are reported in Appendix \ref{app:distance_lift_hardbc}. For the three case studies, the distance function vanishes identically on the domain boundary and reaches its maximum in the interior, so that the network contribution $D(\bm{x})\!\cdot\!\mathcal{N}$ is automatically zero on the boundary, while the lifting function $G(\bm{x})$ reproduces the prescribed boundary arch profile.

\begin{figure*}[H]
  \centering
  \includegraphics[width=\textwidth]{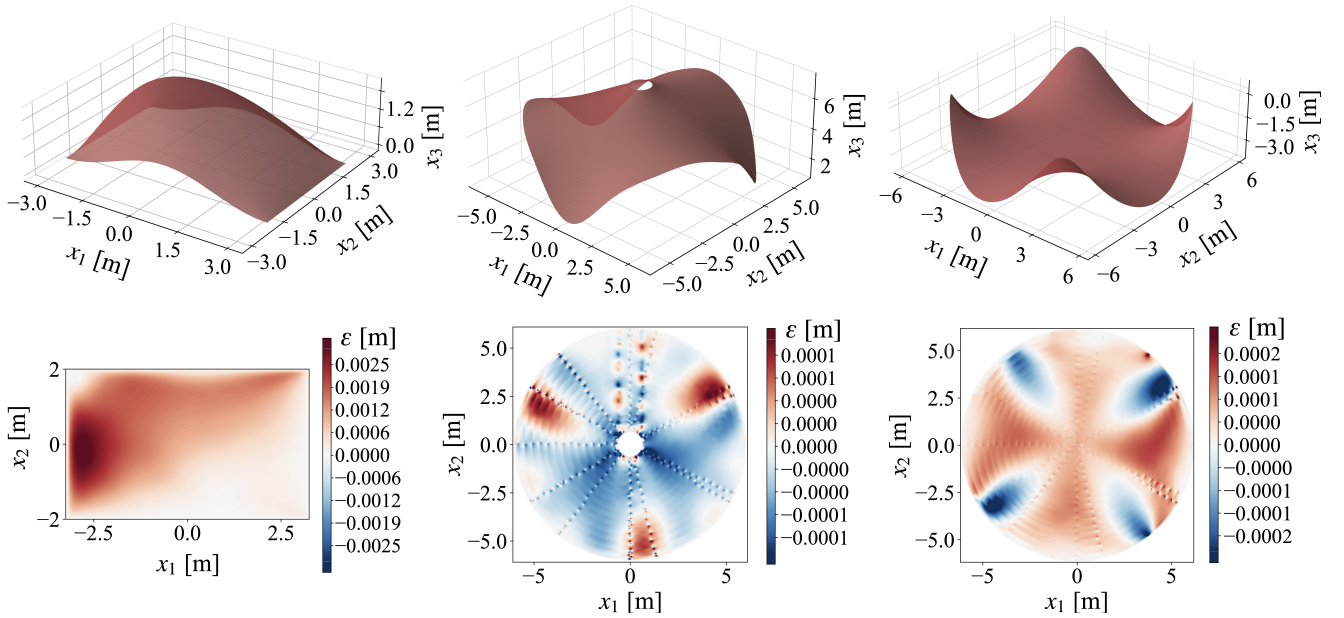}
  \caption{The top row compares the membrane surfaces obtained with the Hard-BC PINN solution (red) and with FEniCSx (gray). The bottom row shows the pointwise difference between the two solutions, computed as $f_{\mathrm{PINN}}-f_{\mathrm{FEniCSx}}$. From left to right, the columns correspond to the rectangular domain, the three-legged domain, and the four-legged domain.}
  \label{fig:diff_hard}
\end{figure*}

The resulting membrane surfaces are compared with the FEniCSx reference membrane in \Cref{fig:diff_hard}. As in the soft-BC case, the PINN predictions (red) and the reference membrane (gray) overlap. The pointwise difference shows that residual error is distributed smoothly over the interior and does not exhibit the boundary-concentrated error observed in \Cref{fig:diff_soft}.

The RMSE between the two solutions is markedly lower than for the soft-BC approach: $1.26\times10^{-3}$\,m for the rectangular domain, $6.65\times10^{-5}$\,m for the three-legged domain, and $7.29\times10^{-5}$\,m for the four-legged domain. The corresponding relative $L^{2}$ errors are $8.8\times10^{-2}\,\%$ for the rectangular domain, $2\times10^{-3}\,\%$ for the three-legged domain, and $3\times10^{-3}\,\%$ for the four-legged domain. The maximum absolute error is reduced to $3.14\times10^{-3}$\,m for the rectangular domain, $3.94\times10^{-4}$\,m for the three-legged domain, and $4.32\times10^{-4}$\,m for the four-legged domain. The errors are summarized in \Cref{tab:results}. The hard-BC PINN performs particularly well near the domain boundary, where the distance function ensures exact satisfaction of the Dirichlet data by construction. In contrast with the soft-BC formulation, the error distribution does not appear to be strongly influenced by the location of the applied load. 

\begin{table*}[H]
  \centering
  \caption{Summary of the accuracy of the PINN solutions against the FEniCSx reference for the three case studies.}
  \label{tab:results}
  \footnotesize
  \begin{tabular}{@{}llcccc@{}}
    \toprule
    Case & Method
      & RMSE [m]
      & Rel $L^{2}$ [\%]
      & Max $|\Delta|$ [m] \\
    \midrule
    \multirow{2}{*}{Rectangular}
      & Soft-BC ($5\!\times\!128$)
        & $1.45\times10^{-3}$
        & 0.116
        & $8.47\times10^{-3}$ \\
      & Hard-BC ($4\!\times\!256$)
        & $1.11\times10^{-3}$
        & 0.088
        & $3.14\times10^{-3}$ \\
    \midrule
    \multirow{2}{*}{Three-legged}
      & Soft-BC ($5\!\times\!128$)
        & $3.82\times10^{-3}$
        & 0.097
        & $3.38\times10^{-2}$ \\
      & Hard-BC ($4\!\times\!256$)
        & $6.65\times10^{-5}$
        & 0.002
        & $3.94\times10^{-4}$ \\
    \midrule
    \multirow{2}{*}{Four-legged}
      & Soft-BC ($5\!\times\!128$)
        & $1.00\times10^{-3}$
        & 0.038
        & $9.69\times10^{-3}$ \\
      & Hard-BC ($4\!\times\!256$)
        & $7.29\times10^{-5}$
        & 0.003
        & $4.32\times10^{-4}$ \\
    \bottomrule
  \end{tabular}
\end{table*}

\subsection{PDE residual validation}
\label{sec:results:pde_validation}

\begin{figure*}[H]
  \centering
  \includegraphics[width=\textwidth]{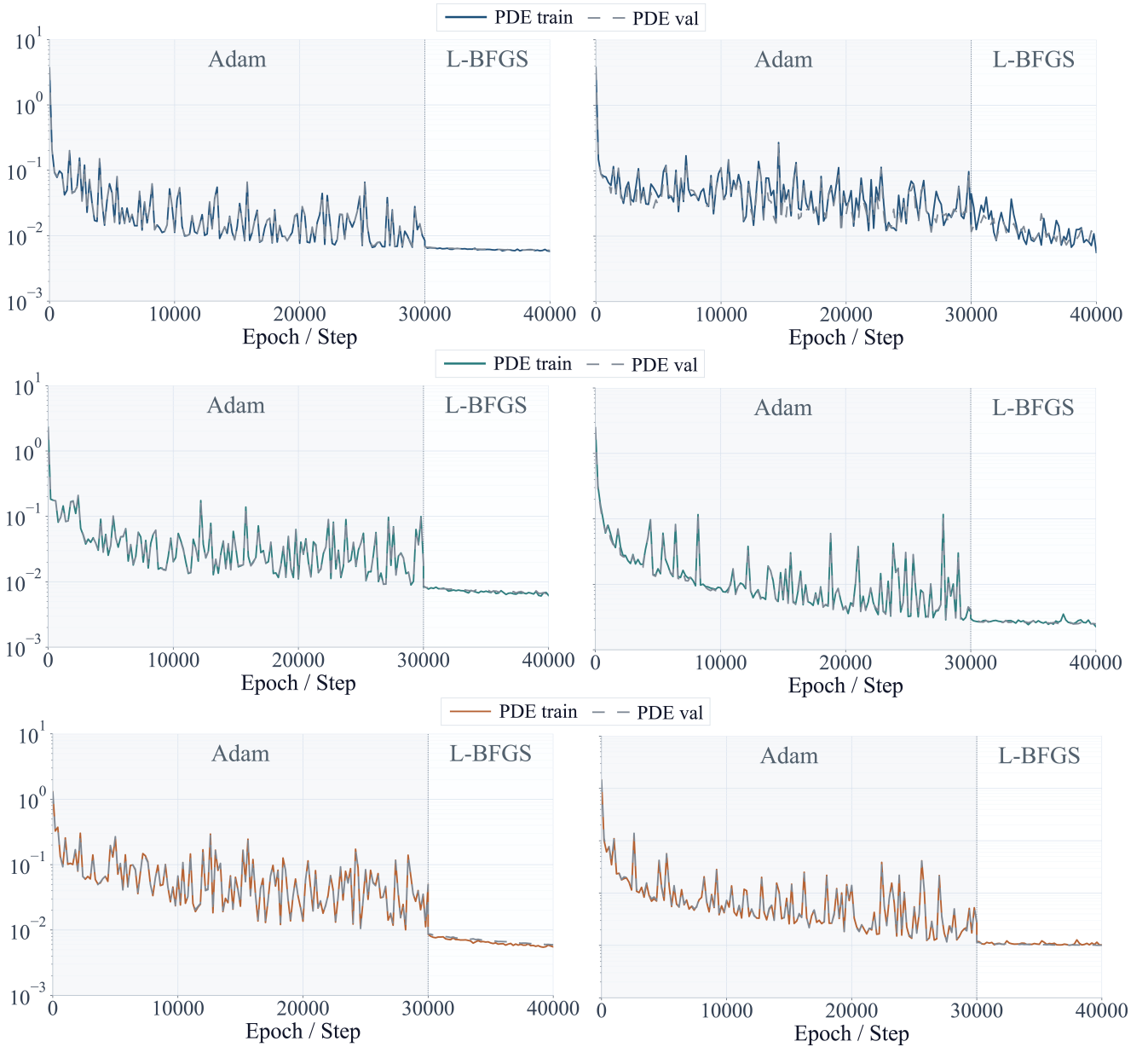}
  \caption{Convergence of the RMSE of the PDE residual on the training collocation point set and
    on the fixed validation point set for the three case studies. The rows correspond, from top to bottom, to the rectangular domain, the three-legged domain, and the four-legged domain. The soft-BC and hard-BC formulations are shown in the left and right panels, respectively. In each panel, solid lines denote the training PDE RMSE and dashed lines denote the validation PDE RMSE. The vertical dashed line marks the transition from Adam training (30\,000 epochs) to L-BFGS refinement (10\,000 steps).}
  \label{fig:pde_convergence}
\end{figure*}

Monitoring the PDE residual during training provides a verification that the predicted membrane retains its physical meaning. Since the membrane shape is defined as the solution of the governing equilibrium equation, a membrane that matches the reference surface but does not satisfy the PDE throughout the domain would not represent a mechanically admissible membrane. The evolution of the RMSE of the PDE residual during training for the three case studies and for both PINN formulations is shown in \Cref{fig:pde_convergence}. 

The training curve is computed on the interior collocation points used by the optimizer, whereas the validation curve is evaluated on an independently sampled set of interior points that is never used for parameter updates. Since both point sets are drawn from the same domain and loading conditions, this validation is not intended to provide an independent accuracy measure with respect to the reference FEniCSx solution. Instead, it is used to assess whether the reduction of the PDE residual achieved during training generalizes to unseen interior points of the same physical problem.

For all case studies and both PINN formulations, the training and validation curves remain in close agreement throughout the optimization. No persistent difference is observed in either the Adam or the L-BFGS stage. This provides an indicator of solution consistency, as the admissibility of the membrane surface requires that the governing equation is satisfied throughout the interior and not only at the points used during optimization. In all cases, the largest reduction in PDE RMSE occurs during the Adam stage, while the subsequent L-BFGS stage continues to reduce the residual more gradually. 

The final validation PDE RMSE values are $8.7\!\times\!10^{-3}$, $2.5\!\times\!10^{-3}$ and $1.0\!\times\!10^{-3}$ for the rectangular, three-legged and four-legged domains under the hard-BC formulation, and of comparable order for the soft-BC formulation. These residuals are several orders of magnitude smaller than the magnitudes of the applied loads reported in \Cref{fig:applied_loads}, indicating that both PINN formulations satisfy the membrane equilibrium equation with good accuracy over the interior domain.

\subsection{Convergence to the reference solution}
\label{sec:results:convergence}

\begin{figure*}[H]
  \centering
  \includegraphics[width=\textwidth]{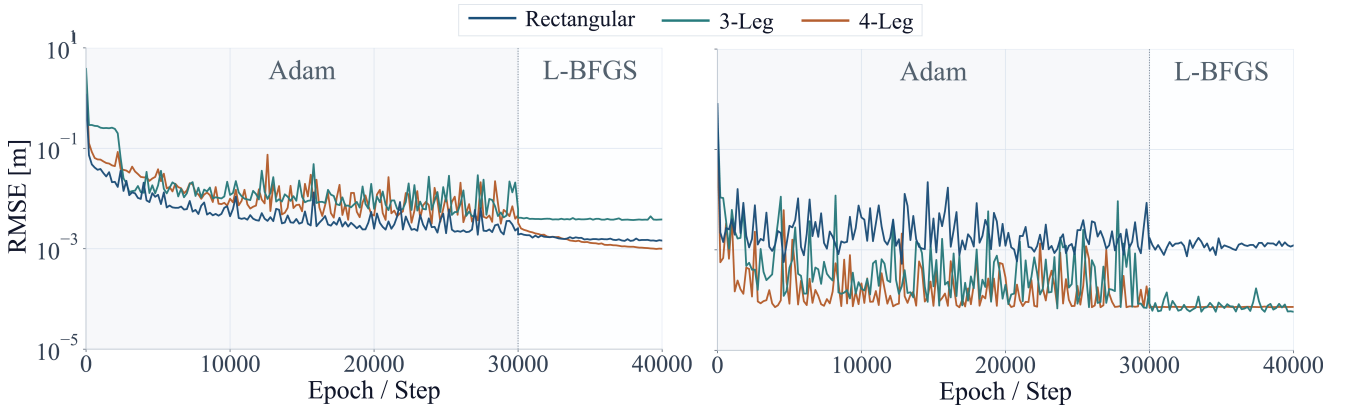}
  \caption{RMSE convergence of the PINN solutions relative to the FEniCSx reference for the three case studies. The soft-BC and hard-BC formulations are shown in the left and right panels, respectively. The dashed vertical line marks the transition from Adam training (30\,000 epochs) to L-BFGS refinement (10\,000 steps). The curves correspond to the rectangular, three-legged, and four-legged domains.}
  \label{fig:convergence}
\end{figure*}

The RMSE with respect to the FEniCSx reference is not included in the loss function of either PINN formulation, but it provides a useful measure of how rapidly the predicted membrane approaches the reference solution during training. \Cref{fig:convergence} shows the evolution of the RMSE between the solutions of the two PINN formulations and the FEniCSx reference during training for all three case studies.

For the soft-BC formulation, the error decreases rapidly during the first few hundred Adam epochs in all three cases. Thereafter, the convergence differs across the domains. After the initial drop, the rectangular and four-legged domains exhibit a more regular and monotonic decrease over most of the Adam stage. By contrast, the three-legged shows an earlier temporary flattening, followed by a sharp reduction at roughly 2\,000 epochs. Overall, the Adam-stage convergence tends to flatten after roughly 20\,000 epochs for all three case studies. After the transition to L-BFGS, all three cases undergo a further refinement, with the most visible additional decrease occurring in the four-legged domain. Overall, the soft-BC results indicate that most of the accuracy is already achieved during the Adam stage, while L-BFGS provides a final improvement.

The hard-BC formulation converges more rapidly. For all three case studies, the RMSE drops sharply within the first few hundred Adam epochs, after which it is already close to its final values for the rectangular and four-legged domains. The three-legged domain requires roughly 3\,000 epochs to approach a comparable plateau. This faster reduction in reference-membrane RMSE is consistent with the exact imposition of the Dirichlet boundary conditions, which removes the need to balance a boundary penalty against the PDE residual and leaves a single objective to drive the optimization. Although the interior PDE residual decreases for the entire Adam stage (see \Cref{sec:results:pde_validation}), the reference-membrane RMSE shows only limited additional improvement.

While both architectures were trained over a deliberately extensive number of epochs, the PINNs attained accurate solutions already within the first few thousand Adam epochs. This early convergence is quantified in \Cref{tab:early_convergence}, which reports the RMSE after 5\,000 Adam epochs, without any L-BFGS refinement. At this stage, the hard-BC formulation already attains a relative error below $0.1\,\%$, while the soft-BC formulation provides approximations with errors ranging from about $0.3\,\%$ to $0.8\,\%$. It should be noted that the reported convergence curves are specific to the PINN configurations adopted in this work, including the selected hyperparameters, network architectures, and numbers of collocation points. The proposed PINN formulations were calibrated primarily to achieve the best possible accuracy, rather than to minimize the number of training epochs. Different training choices could therefore modify the convergence rate and lead either to faster or to slower convergence.

\begin{table}[H]
  \centering
  \caption{Summary of the accuracy of the PINN solutions against the FEniCSx reference for the three case studies after 5\,000 Adam-only epochs (no L-BFGS refinement).}
  \label{tab:early_convergence}
  \begin{tabular}{@{}llcc@{}}
    \toprule
    Case & Method
      & RMSE [m]
      & Rel $L^{2}$ [\%] \\
    \midrule
    \multirow{2}{*}{Rectangular}
      & Soft-BC ($4\!\times\!128$)
        & $7.42\times10^{-3}$ & 0.593 \\
      & Hard-BC ($3\!\times\!256$)
        & $1.26\times10^{-3}$ & 0.100 \\
    \midrule
    \multirow{2}{*}{Three-legged}
      & Soft-BC ($4\!\times\!128$)
        & $1.18\times10^{-2}$ & 0.300 \\
      & Hard-BC ($3\!\times\!256$)
        & $1.40\times10^{-4}$ & 0.003 \\
    \midrule
    \multirow{2}{*}{Four-legged}
      & Soft-BC ($4\!\times\!128$)
        & $2.03\times10^{-2}$ & 0.771 \\
      & Hard-BC ($3\!\times\!256$)
        & $8.66\times10^{-5}$ & 0.003 \\
    \bottomrule
  \end{tabular}
\end{table}

\subsection{Discussion}
\label{sec:results:discussion}

The results presented in \Cref{sec:results:soft,sec:results:hard} show that both PINN formulations solve the MEA governing PDE with high accuracy in all three case studies. The comparison highlights differences in error distribution and in overall accuracy. In general, the hard-BC formulation provides smaller errors than the soft-BC formulation, with the most evident improvement occurring near the boundary. At the same time, the extent of this improvement depends on how effectively the hard-BC ansatz can be constructed for the prescribed domain and boundary conditions. In the three-legged and four-legged domains, the hard-BC formulation yields substantially lower RMSE, relative \(L^2\) error, and maximum absolute error than the soft-BC formulation. In the rectangular case, however, the benefit is more limited: although the hard-BC formulation still yields smaller values for all the error metrics, the overall performance remains close to that of the soft-BC solution. This suggests that, while exact boundary enforcement is beneficial in principle, its practical advantage also depends on the quality of the chosen distance and lift functions and on the optimization behavior induced by the corresponding hard-BC parameterization.

Nevertheless, the soft-BC formulation remains a practically relevant option. Its lower accuracy compared with the hard-BC formulation does not prevent it from providing structurally meaningful solutions. In the present study, the relative $L^2$ error remains below $0.12\,\%$ for all three case studies, indicating that the predicted membranes remain very close to the reference solutions. This level of accuracy is often sufficient in structural applications where an exact pointwise match of the prescribed boundary data is not required. In form-finding problems, for example, the boundary profile may represent a preliminary design input rather than a strict geometric constraint, and a limited relaxation of the boundary condition may be acceptable provided that the resulting membrane remains close to the target geometry and satisfies equilibrium. A similar argument applies to the limit-analysis assessment of masonry vaults \cite{heyman1977equilibrium, pingaro2025numerical}, where the relevant structural requirement is that an equilibrium thrust membrane lies within the thickness of the vault. In such cases, the obtained error levels for the soft-BC formulation do not compromise the structural integrity.

A further advantage of the soft-BC approach is its simpler implementation on new geometries. In the hard-BC formulation, the construction of a suitable distance function and lifting function requires additional analytical or numerical preprocessing, which may become non-trivial for complex domains or boundary descriptions. By contrast, the soft-BC formulation is defined directly from the governing PDE and the boundary penalty term, without the need for a precomputed lift. This simpler setup can be advantageous in exploratory design studies, where many candidate geometries must be evaluated quickly and where a moderate reduction in accuracy may be acceptable in exchange for a more direct formulation.

\section{Conclusions}
\label{sec:conclusions}

This paper presented the use of PINNs to solve the governing PDE of the MEA for the form-finding of membrane surfaces under prescribed stress states and loading conditions. The study considered two formulations: a soft-BC approach, in which the Dirichlet boundary conditions are enforced through a penalty term in the loss function, and a hard-BC approach, in which the boundary conditions are satisfied exactly by construction through a distance and a lifting function. Both formulations were assessed against membranes obtained using a PDE solver based on the FEM on three case studies with different geometrical complexity.

The PINN formulations solved the MEA governing PDE with high accuracy. For all three case studies, the predicted membrane surfaces are in close agreement with the finite-element reference solutions. The soft-BC formulation achieves relative $L^2$ errors below $0.1\,\%$ in all cases, which indicates that it already provides structurally meaningful solutions for the present problem. The hard-BC formulation consistently yields lower RMSE, lower relative $L^2$ error, and smaller maximum absolute error. Its main advantage lies in the exact satisfaction of the prescribed boundary conditions, which removes the boundary-localized errors observed in the soft-BC solution and leads to a smoother and smaller residual error over the interior of the domain. On the other hand, the soft-BC formulation remains attractive because of its simpler implementation, since it can be defined directly from the governing PDE and the boundary penalty term without requiring the construction of a distance function and a lifting function. The hard-BC model converges faster as it only employs the PDE loss terms. 

From a structural point of view, the hard-BC formulation is the preferred option when the prescribed boundary profile must be satisfied exactly and when the highest possible accuracy is required. The soft-BC formulation remains a relevant alternative in exploratory design and assessment studies, where moderate relaxation of the boundary data is acceptable and a simpler setup is advantageous.

This work therefore shows that PINNs constitute a viable mesh-free framework for solving the MEA governing equation under combined loading conditions. Future research may extend the approach to richer ASF parameterizations, and its integration into broader form-finding and optimization procedures.

\section*{Acknowledgments}

This work was supported by NSF under grant number OAC-2118201.

\printcredits

\newpage
\appendix

\section{Airy stress functions, principal stresses, and principal directions}
\label{app:airy_stresses}

\begin{figure*}[H]
  \centering
  \includegraphics[width=\textwidth]{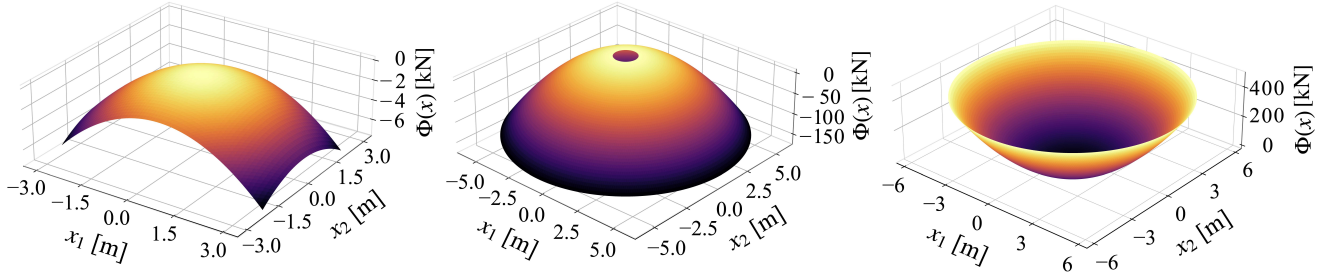}
  \caption{ASF surfaces associated with the three case studies. From left to right, the columns correspond to the rectangular domain, the three-legged domain, and the four-legged domain.}
  \label{fig:airy_stress}
\end{figure*}

\Cref{fig:airy_stress} shows the ASF $\varPhi(x_{1},x_{2})$ associated with the selected stress parameters for the three case studies introduced in \Cref{sec:results:case_studies}. The ASFs are calculated using the parameters reported in \Cref{tab:case_params}, with $c_0=c_1=c_2=0$ so that only the quadratic component associated with the constant Hessian is retained.
Because the adopted parameterization has constant second derivatives, the ASFs are quadratic surfaces over the plan domain. In the rectangular and three-legged cases, the ASF is concave, consistently with the prescribed compression-only stress state. In the four-legged case, the ASF is convex, consistently with the tension-only formulation.

\begin{figure*}[H]
  \centering
  \includegraphics[width=\textwidth]{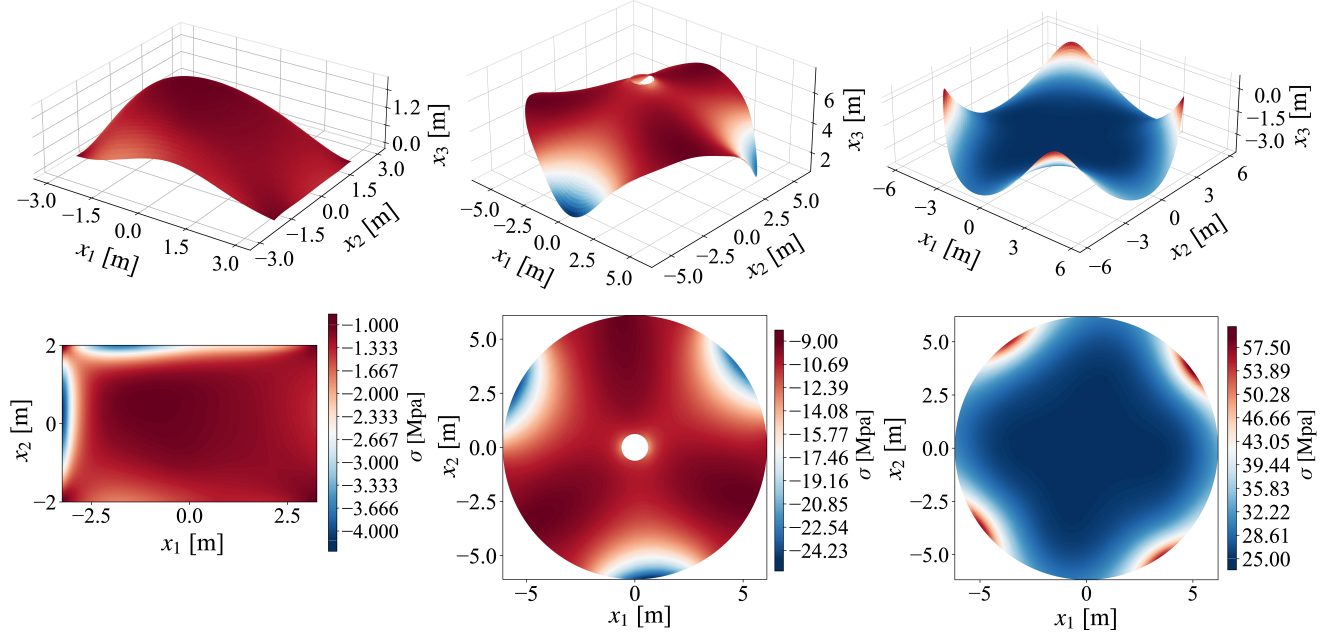}
  \caption{Maximum principal stress $\sigma_{1}$ for the three case studies, shown on the three-dimensional membrane surface (top row) and in plan view (bottom row). From left to right, the columns correspond to the rectangular domain, the three-legged domain, and the four-legged domain}
  \label{fig:max_stresses}
\end{figure*}

\begin{figure*}[H]
  \centering
  \includegraphics[width=\textwidth]{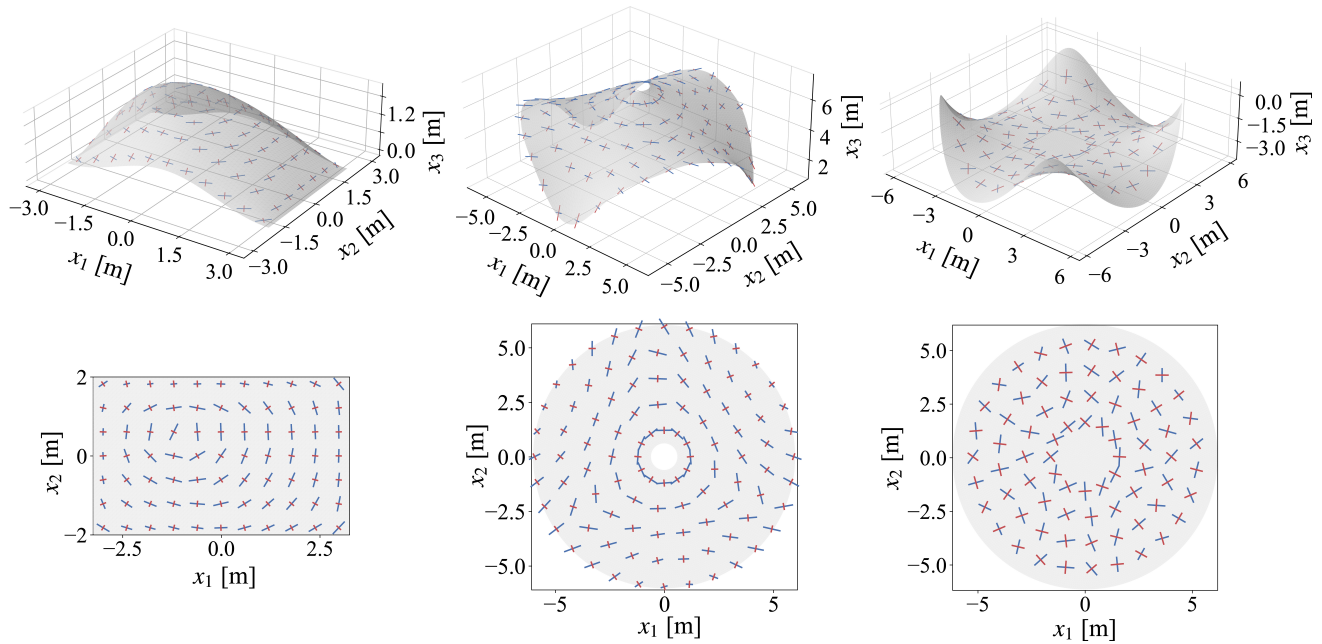}
  \caption{Principal membrane stress directions shown on the three-dimensional membrane surfaces (top row) and in plan view (bottom row). At each sampled point, the blue and red segments indicate the directions associated with the two principal stresses, $\sigma_1$ and $\sigma_2$, respectively. From left to right, the columns correspond to the rectangular domain, the three-legged domain, and the four-legged domain.}
  \label{fig:principal_directions}
\end{figure*}

\Cref{fig:max_stresses} shows the maximum principal membrane stress $\sigma_{1}$ on both the three-dimensional membrane surface and its plan projection. In the rectangular and three-legged cases, $\sigma_{1}$ remains negative over the whole domain, confirming that the membrane stress state is compression-only. In the four-legged case, $\sigma_{1}$ remains positive throughout the domain, consistently with the prescribed tension-only stress state. The principal membrane directions are shown in \Cref{fig:principal_directions} on both the membrane surface and in plan view.

\section{Distance and lift functions for the hard-BC formulation}
\label{app:distance_lift_hardbc}

For completeness, the explicit expressions of the distance function $D(\bm{x})$ and the lift function $G(\bm{x})$ used in the hard-BC ansatz \cref{eq:hard_ansatz} for the three case studies are here reported. In all cases, $D(\bm{x})$ is chosen so that $D=0$ on $\partial\Omega$ and $D>0$ in the interior, while $G(\bm{x})$ satisfies the prescribed Dirichlet boundary condition and is harmonic in the domain.

\subsection{Rectangular domain}
\label{app:distance_lift_rect}

For the rectangular domain of span $l$ and width $b$, the distance function is taken as
\begin{equation}
  \label{eq:app_distance_rect}
  D(\bm{x}) =
  \Bigl(1 - \bigl(\tfrac{2x_{1}}{l}\bigr)^{2}\Bigr)
  \Bigl(1 - \bigl(\tfrac{2x_{2}}{b}\bigr)^{2}\Bigr),
\end{equation}
which vanishes on the four edges of the rectangle and attains its maximum at the center.

The lift function is prescribed analytically as a raised-cosine arch:
\begin{equation}
  \label{eq:app_lift_rect}
  G(x_{1}) =
  \tfrac{1}{2}\,h_{\mathrm{arch}}
  \Bigl(1 + \cos\!\bigl(\pi x_{1}/(l/2)\bigr)\Bigr),
\end{equation}
where $h_{\mathrm{arch}}$ denotes the boundary arch height.

\subsection{Four-legged domain}
\label{app:distance_lift_4legs}

For the four-legged case, the plan domain is embedded in a disk of radius $R$, and the distance function is chosen as
\begin{equation}
  \label{eq:app_distance_4legs}
  D(\bm{x}) = 1 - \frac{x_{1}^{2}+x_{2}^{2}}{R^{2}}.
\end{equation}

Let the prescribed boundary profile on the outer circle $r=R$ be represented by the truncated Fourier expansion
\begin{equation}
  \label{eq:app_bc_4legs}
  g(\theta) =
  a_{0} + \sum_{k=1}^{K}
  \bigl(a_{k}\cos(k\theta) + b_{k}\sin(k\theta)\bigr),
\end{equation}
where $r=\sqrt{x_{1}^{2}+x_{2}^{2}}$ and $\theta$ is the polar angle. The corresponding harmonic lift on the disk is
\begin{equation}
  \label{eq:app_lift_4legs}
  G(r,\theta) =
  a_{0} + \sum_{k=1}^{K}
  \Bigl(\frac{r}{R}\Bigr)^{k}
  \bigl(a_{k}\cos(k\theta) + b_{k}\sin(k\theta)\bigr).
\end{equation}

\subsection{Three-legged annular domain}
\label{app:distance_lift_3legs}

For the three-legged annular case, with inner radius $R_{\mathrm{in}}$ and outer radius $R_{\mathrm{out}}$, the distance function is defined as
\begin{equation}
  \label{eq:app_distance_3legs}
  D(\bm{x}) =
  \frac{(r - R_{\mathrm{in}})(R_{\mathrm{out}} - r)}{D_{\max}},
  \qquad
  D_{\max} =
  \frac{(R_{\mathrm{out}} - R_{\mathrm{in}})^{2}}{4},
\end{equation}
with $r=\sqrt{x_{1}^{2}+x_{2}^{2}}$. This normalized product vanishes on both the inner and outer boundaries.

Let the prescribed boundary conditions on the outer and inner circles be written as
\begin{equation}
  \label{eq:app_bc_3legs_outer}
  g_{\mathrm{out}}(\theta) =
  a_{0}^{\mathrm{out}} + \sum_{k=1}^{K}
  \bigl(a_{k}^{\mathrm{out}}\cos(k\theta) + b_{k}^{\mathrm{out}}\sin(k\theta)\bigr),
\end{equation}
\begin{equation}
  \label{eq:app_bc_3legs_inner}
  g_{\mathrm{in}}(\theta) =
  a_{0}^{\mathrm{in}} + \sum_{k=1}^{K}
  \bigl(a_{k}^{\mathrm{in}}\cos(k\theta) + b_{k}^{\mathrm{in}}\sin(k\theta)\bigr).
\end{equation}

The harmonic lift on the annulus then takes the form
\begin{equation}
\begin{aligned}
  \label{eq:app_lift_3legs}
  G(r,\theta) 
  &= A_{0} + B_{0}\log r\\
  &+ \sum_{k=1}^{K}
  \Bigl[
    \bigl(A_{k}r^{k} + B_{k}r^{-k}\bigr)\cos(k\theta)\\
    &+
    \bigl(C_{k}r^{k} + D_{k}r^{-k}\bigr)\sin(k\theta)
  \Bigr].
\end{aligned}
\end{equation}

The coefficients are determined by matching the inner and outer boundary data. For the axisymmetric term,
\begin{equation}
  \label{eq:app_lift_3legs_k0}
  A_{0} + B_{0}\log R_{\mathrm{out}} = a_{0}^{\mathrm{out}},
  \qquad
  A_{0} + B_{0}\log R_{\mathrm{in}} = a_{0}^{\mathrm{in}},
\end{equation}
while for each harmonic order $k=1,\dots,K$,
\begin{equation}
  \label{eq:app_lift_3legs_cos}
  A_{k}R_{\mathrm{out}}^{k} + B_{k}R_{\mathrm{out}}^{-k} = a_{k}^{\mathrm{out}},
  \qquad
  A_{k}R_{\mathrm{in}}^{k} + B_{k}R_{\mathrm{in}}^{-k} = a_{k}^{\mathrm{in}},
\end{equation}
\begin{equation}
  \label{eq:app_lift_3legs_sin}
  C_{k}R_{\mathrm{out}}^{k} + D_{k}R_{\mathrm{out}}^{-k} = b_{k}^{\mathrm{out}},
  \qquad
  C_{k}R_{\mathrm{in}}^{k} + D_{k}R_{\mathrm{in}}^{-k} = b_{k}^{\mathrm{in}}.
\end{equation}





\begin{thebibliography}{10}
\expandafter\ifx\csname url\endcsname\relax
  \def\url#1{\texttt{#1}}\fi
\expandafter\ifx\csname urlprefix\endcsname\relax\def\urlprefix{URL }\fi
\expandafter\ifx\csname href\endcsname\relax
  \def\href#1#2{#2} \def\path#1{#1}\fi

\bibitem{heyman1966stone}
J.~Heyman, The stone skeleton, International Journal of Solids and Structures 2~(2) (1966) 249--279.
\newblock \href {https://doi.org/10.1016/0020-7683(66)90018-7} {\path{doi:10.1016/0020-7683(66)90018-7}}.

\bibitem{heyman1995stone}
J.~Heyman, The stone skeleton: structural engineering of masonry structures (1995).

\bibitem{heyman1977equilibrium}
J.~Heyman, Equilibrium of shell structures, Oxford: Clarendon Press, 1977.

\bibitem{adriaenssens2014shell}
S.~Adriaenssens, P.~Block, D.~Veenendaal, C.~Williams, Shell structures for architecture: form finding and optimization, Routledge, 2014.

\bibitem{melchiorre2025form}
J.~Melchiorre, A.~M. Bertetto, S.~Adriaenssens, G.~C. Marano, Form-finding and metaheuristic multiobjective optimization methodology for sustainable gridshells with reduced construction complexity and waste, Automation in Construction 177 (2025).

\bibitem{bertetto2024improved}
A.~M. Bertetto, J.~Melchiorre, G.~C. Marano, Improved multi-body rope approach for free-form gridshell structures using equal-length element strategy, Automation in Construction 161 (2024) 105340.
\newblock \href {https://doi.org/10.1016/j.autcon.2024.105340} {\path{doi:10.1016/j.autcon.2024.105340}}.

\bibitem{olivieri2021parametric}
C.~Olivieri, M.~Angelillo, A.~Gesualdo, A.~Iannuzzo, A.~Fortunato, Parametric design of purely compressed shells, Mechanics of Materials 155 (2021).
\newblock \href {https://doi.org/10.1016/j.mechmat.2021.103782} {\path{doi:10.1016/j.mechmat.2021.103782}}.

\bibitem{angelillo2013singular}
M.~Angelillo, E.~Babilio, A.~Fortunato, Singular stress fields for masonry-like vaults, Continuum Mechanics and Thermodynamics 25 (2013) 423--441.
\newblock \href {https://doi.org/10.1007/s00161-012-0270-9} {\path{doi:10.1007/s00161-012-0270-9}}.

\bibitem{pucher1934spannungszustand}
A.~Pucher, Uber den {S}pannungszustand in gekr{\"u}mmten {F}l{\"a}chen, Beton und Eisen 33~(19) (1934) 298--304.

\bibitem{angelillo2004equilibrium}
M.~Angelillo, A.~Fortunato, Equilibrium of masonry vaults, in: Novel approaches in civil engineering, Springer, 2004, pp. 105--111.

\bibitem{olivieri2025seismic}
C.~Olivieri, S.~Cocking, F.~Fabbrocino, A.~Iannuzzo, L.~Placidi, S.~Adriaenssens, Seismic capacity of purely compressed shells based on {A}iry stress function, Continuum Mechanics and Thermodynamics 37~(21) (2025).
\newblock \href {https://doi.org/10.1007/s00161-024-01350-z} {\path{doi:10.1007/s00161-024-01350-z}}.

\bibitem{naghdi1973theory}
P.~M. Naghdi, The theory of shells and plates, in: Linear theories of elasticity and thermoelasticity: linear and nonlinear theories of rods, plates, and shells, Springer, 1973, pp. 425--640.

\bibitem{olivieri2023formerly}
C.~Olivieri, {FORMERLY-Math: Constrained form-finding through membrane equilibrium analysis in Mathematica}, Software Impacts 16 (2023) 100512.
\newblock \href {https://doi.org/10.1016/j.simpa.2023.100512} {\path{doi:10.1016/j.simpa.2023.100512}}.

\bibitem{baratta_2023_10447666}
I.~A. Baratta, J.~P. Dean, J.~S. Dokken, M.~Habera, J.~S. Hale, C.~N. Richardson, M.~E. Rognes, M.~W. Scroggs, N.~Sime, G.~N. Wells, Dolfinx: The next generation fenics problem solving environment (Dec. 2023).
\newblock \href {https://doi.org/10.5281/zenodo.10447666} {\path{doi:10.5281/zenodo.10447666}}.

\bibitem{page1978finite}
A.~W. Page, Finite element model for masonry, Journal of the Structural Division 104~(8) (1978) 1267--1285.

\bibitem{pingaro2026simple}
N.~Pingaro, G.~Milani, Simple interface element equipped with thickness for the non-linear static heterogeneous analysis of masonry walls in-plane loaded: Implementation and validation, Engineering Structures 351 (2026) 122056.

\bibitem{chapelle2011finite}
D.~Chapelle, K.-J. Bathe, et~al., The finite element analysis of shells: fundamentals, Vol.~1, Springer, 2011.

\bibitem{Bastek2023}
J.-H. Bastek, D.~M. Kochmann, Physics-informed neural networks for shell structures, European Journal of Mechanics -- A/Solids 97 (2023) 104849.

\bibitem{raissi2019physics}
M.~Raissi, P.~Perdikaris, G.~E. Karniadakis, {Physics-informed neural networks: A deep learning framework for solving forward and inverse problems involving nonlinear partial differential equations}, Journal of Computational physics 378 (2019) 686--707.
\newblock \href {https://doi.org/10.1016/j.jcp.2018.10.045} {\path{doi:10.1016/j.jcp.2018.10.045}}.

\bibitem{son2023novel}
S.~Son, H.~Lee, D.~Jeong, K.-Y. Oh, K.~H. Sun, A novel physics-informed neural network for modeling electromagnetism of a permanent magnet synchronous motor, Advanced Engineering Informatics 57 (2023) 102035.

\bibitem{zhou2025physics}
Z.~Zhou, B.~Johns, Y.~Fang, Y.~Bai, E.~Abdi, Physics-informed neural network for load sway prediction in travelling autonomous mobile cranes, Advanced Engineering Informatics 65 (2025) 103269.

\bibitem{Cai2021}
S.~Cai, Z.~Wang, S.~Wang, P.~Perdikaris, G.~E. Karniadakis, Physics-informed neural networks for heat transfer problems, Journal of Heat Transfer 143~(6) (2021) 060801.
\newblock \href {https://doi.org/10.1115/1.4050542} {\path{doi:10.1115/1.4050542}}.

\bibitem{Haghighat2021}
E.~Haghighat, M.~Raissi, A.~Moure, H.~Gomez, R.~Juanes, A physics-informed deep learning framework for inversion and surrogate modeling in solid mechanics, Computer Methods in Applied Mechanics and Engineering 379 (2021) 113741.
\newblock \href {https://doi.org/https://doi.org/10.1016/j.cma.2021.113741} {\path{doi:https://doi.org/10.1016/j.cma.2021.113741}}.

\bibitem{wang2021understanding}
S.~Wang, Y.~Teng, P.~Perdikaris, Understanding and mitigating gradient flow pathologies in physics-informed neural networks, SIAM Journal on Scientific Computing 43~(5) (2021) A3055--A3081.

\bibitem{krishnapriyan2021characterizing}
A.~Krishnapriyan, A.~Gholami, S.~Zhe, R.~Kirby, M.~W. Mahoney, Characterizing possible failure modes in physics-informed neural networks, Advances in neural information processing systems 34 (2021) 26548--26560.

\bibitem{Bischof2025}
R.~Bischof, M.~A. Kraus, Multi-objective loss balancing for physics-informed deep learning, Computer Methods in Applied Mechanics and Engineering 439 (2025) 117914.
\newblock \href {https://doi.org/https://doi.org/10.1016/j.cma.2025.117914} {\path{doi:https://doi.org/10.1016/j.cma.2025.117914}}.

\bibitem{kharazmi2019variational}
E.~Kharazmi, Z.~Zhang, G.~E. Karniadakis, Variational physics-informed neural networks for solving partial differential equations, arXiv preprint arXiv:1912.00873 (2019).

\bibitem{jagtap2020locally}
A.~D. Jagtap, K.~Kawaguchi, G.~Em~Karniadakis, Locally adaptive activation functions with slope recovery for deep and physics-informed neural networks, Proceedings of the Royal Society A: Mathematical, Physical and Engineering Sciences 476~(2239) (2020).

\bibitem{Wu2023}
C.~Wu, M.~Zhu, Q.~Tan, Y.~Kartha, L.~Lu, A comprehensive study of non-adaptive and residual-based adaptive sampling for physics-informed neural networks, Computer Methods in Applied Mechanics and Engineering 403 (2023) 115671.
\newblock \href {https://doi.org/https://doi.org/10.1016/j.cma.2022.115671} {\path{doi:https://doi.org/10.1016/j.cma.2022.115671}}.

\bibitem{Lagaris1998}
I.~Lagaris, A.~Likas, D.~Fotiadis, Artificial neural networks for solving ordinary and partial differential equations, IEEE Transactions on Neural Networks 9~(5) (1998) 987--1000.
\newblock \href {https://doi.org/10.1109/72.712178} {\path{doi:10.1109/72.712178}}.

\bibitem{berg2018unified}
J.~Berg, K.~Nystr{\"o}m, A unified deep artificial neural network approach to partial differential equations in complex geometries, Neurocomputing 317 (2018) 28--41.

\bibitem{sukumar2022exact}
N.~Sukumar, A.~Srivastava, Exact imposition of boundary conditions with distance functions in physics-informed deep neural networks, Computer Methods in Applied Mechanics and Engineering 389 (2022) 114333.

\bibitem{hendrycks2016gaussian}
D.~Hendrycks, K.~Gimpel, Gaussian error linear units (gelus), arXiv preprint arXiv:1606.08415 (2016).

\bibitem{goodfellow2016deep}
I.~Goodfellow, Y.~Bengio, A.~Courville, Y.~Bengio, Deep learning, Vol.~1, MIT press Cambridge, 2016.

\bibitem{ramachandran2017searching}
P.~Ramachandran, B.~Zoph, Q.~V. Le, Searching for activation functions, arXiv preprint arXiv:1710.05941 (2017).

\bibitem{elfwing2018sigmoid}
S.~Elfwing, E.~Uchibe, K.~Doya, Sigmoid-weighted linear units for neural network function approximation in reinforcement learning, Neural networks 107 (2018) 3--11.

\bibitem{kingma2014adam}
D.~P. Kingma, J.~Ba, Adam: A method for stochastic optimization, arXiv preprint arXiv:1412.6980 (2014).

\bibitem{liu1989limited}
D.~C. Liu, J.~Nocedal, On the limited memory bfgs method for large scale optimization, Mathematical programming 45~(1) (1989) 503--528.

\bibitem{daw2022rethinking}
A.~Daw, J.~Bu, S.~Wang, P.~Perdikaris, A.~Karpatne, Rethinking the importance of sampling in physics-informed neural networks, arXiv preprint arXiv:2207.02338 (2022).

\bibitem{daw2022mitigating}
A.~Daw, J.~Bu, S.~Wang, P.~Perdikaris, A.~Karpatne, Mitigating propagation failures in physics-informed neural networks using retain-resample-release (r3) sampling, arXiv preprint arXiv:2207.02338 (2022).

\bibitem{olivieri2023continuous}
C.~Olivieri, A.~Iannuzzo, A.~Montanino, F.~L. Perelli, I.~Elia, S.~Adriaenssens, A continuous stress-based form finding approach for compressed membranes, International Journal of Masonry Research and Innovation 9~(5/6) (2024) 585--605.
\newblock \href {https://doi.org/10.1504/IJMRI.2024.10063945} {\path{doi:10.1504/IJMRI.2024.10063945}}.

\bibitem{heydari2019softadapt}
A.~A. Heydari, C.~A. Thompson, A.~Mehmood, Softadapt: Techniques for adaptive loss weighting of neural networks with multi-part loss functions, arXiv preprint arXiv:1912.12355 (2019).

\bibitem{pingaro2025numerical}
N.~Pingaro, M.~Buzzetti, A.~Gandolfi, A numerical strategy to assess the stability of curved masonry structures using a simple nonlinear truss model, Buildings 15~(13) (2025) 2226.

\end{thebibliography}
\end{document}